\documentclass[prd,superscriptaddress,nofootinbib,floatfix,notitlepage]{revtex4-1}
\pdfoutput=1
\usepackage{amsmath,amssymb,epsfig}
\usepackage{hyperref}
\usepackage{natbib,ifthen}
\usepackage{dcolumn} 

\usepackage{mathrsfs}
\usepackage{simplewick}
\usepackage[lofdepth,lotdepth,caption=false]{subfig}
\usepackage{color}
\usepackage{bbold}

\begin{document}

\newcommand{\jcap}{JCAP}
\newcommand{\apjl}{APJL~}
\newcommand{\physrep}{Physics Reports}
\newcommand{\aap}{Astronomy \& Astrophysics}

\def\Btheo{{B_\delta^{\textrm{theo}}}}
\newcommand{\todo}[1]{{\color{blue}{#1}}} 
\newcommand{\changed}[1]{#1} 
\newcommand{\lbar}[1]{\underline{l}_{#1}}
\newcommand{\drm}{\mathrm{d}}
\renewcommand{\d}{\mathrm{d}}
\renewcommand{\rm}[1]{\mathrm{#1}}
\newcommand{\gaensli}[1]{\lq #1\rq$ $}
\newcommand{\bartilde}[1]{\bar{\tilde #1}}
\newcommand{\barti}[1]{\bartilde{#1}}
\newcommand{\ti}{\tilde}
\newcommand{\oforder}[1]{\mathcal{O}(#1)}
\newcommand{\D}{\mathrm{D}}
\renewcommand{\(}{\left(}
\renewcommand{\)}{\right)}
\renewcommand{\[}{\left[}
\renewcommand{\]}{\right]}
\def\<{\left\langle}
\def\>{\right\rangle}
\newcommand{\mycaption}[1]{\caption{\footnotesize{#1}}}
\newcommand{\hattilde}[1]{\hat{\tilde #1}}
\newcommand{\mycite}[1]{[#1]}
\newcommand{\mnras}{Mon.\ Not.\ R.\ Astron.\ Soc.}
\newcommand{\apjs}{Astrophys.\ J.\ Supp.}

\def\uk{{\bf \hat{k}}}
\def\un{{\bf \hat{n}}}
\def\ur{{\bf \hat{r}}}
\def\ux{{\bf \hat{x}}}
\def\bk{{\bf k}}
\def\bn{{\bf n}}
\def\br{{\bf r}}
\def\bx{{\bf x}}
\def\bK{{\bf K}}
\def\by{{\bf y}}
\def\bl{{\bf l}}
\def\bkp{{\bf k^\pr}}
\def\brp{{\bf r^\pr}}

\newcommand{\fixme}[1]{{\textbf{Fixme: #1}}}
\newcommand{\detD}{{\det\!\cld}}
\newcommand{\clh}{\mathcal{H}}
\newcommand{\ud}{{\rm d}}
\renewcommand{\eprint}[1]{\href{http://arxiv.org/abs/#1}{#1}}
\newcommand{\adsurl}[1]{\href{#1}{ADS}}
\newcommand{\ISBN}[1]{\href{http://cosmologist.info/ISBN/#1}{ISBN: #1}}
\newcommand{\vort}{\varpi}
\newcommand\ba{\begin{eqnarray}}
\newcommand\ea{\end{eqnarray}}
\newcommand\be{\begin{equation}}
\newcommand\ee{\end{equation}}
\newcommand\lagrange{{\cal L}}
\newcommand\cll{{\cal L}}
\newcommand\cln{{\cal N}}
\newcommand\clx{{\cal X}}
\newcommand\clz{{\cal Z}}
\newcommand\clv{{\cal V}}
\newcommand\cld{{\cal D}}
\newcommand\clt{{\cal T}}

\newcommand\clo{{\cal O}}
\newcommand{\cla}{{\cal A}}
\newcommand{\clp}{{\cal P}}
\newcommand{\clr}{{\cal R}}
\newcommand{\uD}{{\mathrm{D}}}
\newcommand{\calE}{{\cal E}}
\newcommand{\calB}{{\cal B}}
\newcommand{\curl}{\,\mbox{curl}\,}
\newcommand\del{\nabla}
\newcommand\Tr{{\rm Tr}}
\newcommand\half{{\frac{1}{2}}}
\newcommand\fourth{{1\over 8}}
\newcommand\bibi{\bibitem}
\newcommand{\kf}{\beta}
\newcommand{\kfprod}{\alpha}
\newcommand\calS{{\cal S}}
\renewcommand\H{{\cal H}}
\newcommand\K{{\rm K}}
\newcommand\mK{{\rm mK}}
\newcommand\synch{\text{syn}}
\newcommand\opacity{\tau_c^{-1}}

\newcommand{\Psil}{\Psi_l}
\newcommand{\bsigma}{{\bar{\sigma}}}
\newcommand{\bI}{\bar{I}}
\newcommand{\bq}{\bar{q}}
\newcommand{\bv}{\bar{v}}
\renewcommand\P{{\cal P}}
\newcommand{\numfrac}[2]{{\textstyle \frac{#1}{#2}}}

\newcommand{\la}{\langle}
\newcommand{\ra}{\rangle}
\newcommand{\lla}{\left\langle}
\newcommand{\rra}{\right\rangle}

\newcommand{\Omtot}{\Omega_{\mathrm{tot}}}
\newcommand\xx{\mbox{\boldmath $x$}}
\newcommand{\phpr} {\phi'}
\newcommand{\gam}{\gamma_{ij}}
\newcommand{\sqgam}{\sqrt{\gamma}}
\newcommand{\delk}{\Delta+3{\K}}
\newcommand{\dph}{\delta\phi}
\newcommand{\om} {\Omega}
\newcommand{\dom}{\delta^{(3)}\left(\Omega\right)}
\newcommand{\rar}{\rightarrow}
\newcommand{\Rar}{\Rightarrow}
\newcommand\gsim{ \lower .75ex \hbox{$\sim$} \llap{\raise .27ex \hbox{$>$}} }
\newcommand\lsim{ \lower .75ex \hbox{$\sim$} \llap{\raise .27ex \hbox{$<$}} }
\newcommand\bigdot[1] {\stackrel{\mbox{{\huge .}}}{#1}}
\newcommand\bigddot[1] {\stackrel{\mbox{{\huge ..}}}{#1}}
\newcommand{\Mpc}{\text{Mpc}}
\newcommand{\Al}{{A_l}}
\newcommand{\Bl}{{B_l}}
\newcommand{\eAl}{e^\Al}
\newcommand{\ix}{{(i)}}
\newcommand{\ixp}{{(i+1)}}
\renewcommand{\k}{\beta}
\newcommand{\HD}{\mathrm{D}}

\newcommand{\nonflat}[1]{#1}
\newcommand{\Cgl}{C_{\text{gl}}}
\newcommand{\Cgltwo}{C_{\text{gl},2}}
\newcommand{\He}{{\rm{He}}}
\newcommand{\Mhz}{{\rm MHz}}
\newcommand{\vx}{{\mathbf{x}}}
\newcommand{\ve}{{\mathbf{e}}}
\newcommand{\vv}{{\mathbf{v}}}
\newcommand{\vk}{{\mathbf{k}}}
\newcommand{\vn}{{\mathbf{n}}}
\newcommand{\vPsi}{{\mathbf{\Psi}}}
\newcommand{\vs}{{\mathbf{s}}}
\newcommand{\vH}{{\mathbf{H}}}
\newcommand{\theo}{\mathrm{th}}
\newcommand{\lin}{\mathrm{lin}}
\newcommand{\rec}{\mathrm{rec}}

\newcommand{\vnhat}{{\hat{\mathbf{n}}}}
\newcommand{\vkhat}{{\hat{\mathbf{k}}}}
\newcommand{\taueps}{{\tau_\epsilon}}

\newcommand{\vgrad}{{\mathbf{\nabla}}}
\newcommand{\fbarln}{\bar{f}_{,\ln\epsilon}(\epsilon)}

\newcommand{\secref}[1]{Section \ref{#1}}
\newcommand{\expt}{\mathrm{expt}}
\newcommand{\eq}[1]{(\ref{eq:#1})} 
\newcommand{\eqq}[1]{Eq.~(\ref{eq:#1})} 
\newcommand{\fig}[1]{Fig.~\ref{fig:#1}} 
\renewcommand{\to}{\rightarrow}
\renewcommand{\(}{\left(}
\renewcommand{\)}{\right)}
\renewcommand{\[}{\left[}
\renewcommand{\]}{\right]}
\renewcommand{\vec}[1]{\mathbf{#1}}
\newcommand{\vy}{\vec{y}}
\newcommand{\vz}{\vec{z}}
\newcommand{\vq}{\vec{q}}
\newcommand{\VPsi}{\vec{\Psi}}
\newcommand{\vecv}{\vec{v}}
\newcommand{\vnabla}{\vec{\nabla}}
\newcommand{\vl}{\vec{l}}
\newcommand{\vL}{\vec{L}}
\newcommand{\dl}{\d^2\vl}
\newcommand{\valpha}{\vec{\alpha}}
\renewcommand{\L}{\mathscr{L}}

\newcommand{\abs}[1]{\lvert #1\rvert}

\newcommand{\ul}{\underline{l}}
\newcommand{\yucode}{\textsf{FastPM}~{}}


\thispagestyle{empty}

\title{Eulerian BAO Reconstructions and N-Point Statistics}

\author{Marcel Schmittfull}
\affiliation{Berkeley Center for Cosmological Physics, Department of Physics and Lawrence Berkeley National Laboratory, University of California, Berkeley, CA 94720, USA}

\author{Yu Feng}
\affiliation{Berkeley Center for Cosmological Physics, Department of Physics and Lawrence Berkeley National Laboratory, University of California, Berkeley, CA 94720, USA}

\author{Florian Beutler}
\affiliation{Lawrence Berkeley National Laboratory, 1 Cyclotron Road, Berkeley, CA 94720, USA}

\author{Blake Sherwin}
\affiliation{Berkeley Center for Cosmological Physics, Department of Physics and Lawrence Berkeley National Laboratory, University of California, Berkeley, CA 94720, USA}
\affiliation{Miller Institute for Basic Research in Science, University of California, Berkeley, CA, 94720, USA}

\author{Man Yat Chu}
\affiliation{Berkeley Center for Cosmological Physics, Department of Physics and Lawrence Berkeley National Laboratory, University of California, Berkeley, CA 94720, USA}

\date{\today}

\begin{abstract}
As galaxy surveys begin to measure the imprint of baryonic acoustic oscillations (BAO) on large-scale structure at the sub-percent level, reconstruction techniques that reduce the contamination from nonlinear clustering become increasingly important. Inverting the nonlinear continuity equation, we propose an Eulerian growth-shift reconstruction algorithm that does not require the displacement of any objects, which is needed for the standard Lagrangian BAO reconstruction algorithm. \changed{In real-space DM-only simulations the algorithm yields $95\%$ of the BAO signal-to-noise obtained from standard reconstruction.} The reconstructed power spectrum is obtained by adding specific simple 3- and 4-point statistics to the pre-reconstruction power spectrum, making it very transparent how additional BAO information from higher-point statistics is included in the power spectrum through the reconstruction process. Analytical models of the reconstructed density for the two algorithms agree at second order. Based on similar modeling efforts, we introduce four additional reconstruction algorithms and discuss their performance.
\end{abstract}

\maketitle

\section{Introduction}

Measuring the imprint of baryonic acoustic oscillations (BAO) on large-scale structure is becoming increasingly important for tightening constraints on the $\Lambda$CDM model parameters and probing the nature of dark energy and the geometry of the universe; see e.g.~\cite{1201.2434} for a recent review.  The BAO lead to a bump in the 2-point halo correlation function at separation $r\sim 150\,\mathrm{Mpc}$, and to wiggles in the Fourier space power spectrum.\footnote{This BAO imprint can be understood from a simple physical picture by considering perturbations to an initially point-like overdensity at the origin \cite{0604361}.  Dark matter just gravitationally clusters around the origin.  Baryons and photons are tightly coupled before recombination ($z\gtrsim 1100$) and undergo acoustic oscillations, because infall due to gravitational clustering of baryons is counter-acted by photon pressure.  As the universe cools, baryons and photons decouple, so that photons free-stream and baryons cluster in a spherical shell around the origin whose radius is given by the distance the sound wave traveled between the initial time and the time of decoupling; the BAO scale.  The dark matter in the origin and the baryons in the shell around the origin subsequently cluster gravitationally and ultimately form halos which are separated by the BAO scale.  For realistic initial overdensities, this effect is still present statistically. }   Measuring the BAO scale from this provides a standard ruler for a physical scale at low redshift which is very powerful in breaking parameter degeneracies that would be present when considering the high-redshift Cosmic Microwave Background (CMB) alone (e.g.~\cite{PlanckParams2013, PlanckParams2015, Mehta1202.0092, Anderson1303, Aouburg1411}).   Measuring the distinct large-scale BAO bump in the correlation function or the corresponding wiggles in the power spectrum is also very robust against potential astrophysical and observational systematics. 

Nonlinear gravitational clustering smears the sharp linear BAO bump in the correlation function, mostly due to large-scale bulk flows and cluster formation \cite{0604361}.  In a procedure known as BAO reconstruction these large-scale shifts are reversed to (partially) recover the linear BAO signal \cite{EisensteinRecSims0604362}.  Slightly more generally, the goal of large scale structure reconstruction is to restore information about the linear modes that were corrupted by nonlinear structure formation.  In practice, BAO reconstruction works extraordinarily well and reduces the uncertainty on the BAO scale significantly, e.g.~by a factor of more than $1.5$ for BOSS DR11 \cite{Anderson1312.4877}  (see also \cite{Padmanabhan2012BAORec,AndersonBAODR9_1203,1409.3242,2014MNRAS.441.3524K,Florian1506.03900}).  Reconstruction also reduces a small bias of the BAO scale that is caused by nonlinear clustering.  This has been studied in detail analytically and in simulations \cite{padmanabhan0812,NohPadmanabhanWhite2009,SherwinZaldarriaga2012,TassevRec,SeoBAOShift,Burden1408,Achitouv2015,white1504.03677,Mehta1104_RecBias}.

Although the standard Lagrangian reconstruction algorithm works very well, there are several motivations to explore new reconstruction algorithms, e.g.
\begin{itemize}
\item to provide more insight where additional BAO information comes from,
\item derive algorithms more rigorously,
\item improve the methodology by requiring fewer model assumptions for data processing,
\item increase BAO information further.
\end{itemize}
Motivated by this, we propose a number of new reconstruction algorithms.  Inspired by the Lagrangian and Eulerian perspectives of fluid dynamics, where either individual particle positions or fields like the mass density are regarded as the fundamental dynamic variable, 
we separate these reconstruction algorithms into two broad, disjoint categories: \textbf{Lagrangian reconstruction algorithms}, which involve displacing individual objects at some point in the reconstruction process, and \textbf{Eulerian reconstruction algorithms}, which do not involve displacing individual objects at any point in the reconstruction process, e.g.~by operating only on density fields.  In this notion `mixed' algorithms like the standard BAO reconstruction, which moves individual objects by a displacement field which is calculated from the linearized continuity equation for the mass density field, are still categorized as Lagrangian because they move individual objects at some step of the algorithm.  A very simple (though not particularly useful) example for an Eulerian reconstruction algorithm is to suppress nonlinear peaks of the mass density by subtracting the squared density from the original density, i.e.~$\delta_\mathrm{rec}(\vx)=\delta(\vx)-\delta^2(\vx)$, which does not involve displacing any objects.

Although we cannot address all motivational bullet points above at once, we use them as guidelines for the Eulerian reconstructions introduced in this paper.  Addressing the first point, we find that the power spectrum after our Eulerian reconstructions can be expressed as a sum of the unreconstructed power spectrum and specific simple 3- and 4-point statistics of the unreconstructed density.  This makes it very transparent and explicit how the additional information on the BAO scale due to reconstruction comes from specific higher order N-point statistics of the unreconstructed density.  Lagrangian reconstructions must also use higher order N-point information implicitly, because phase information of the density is used so that a reconstructed power spectrum realization cannot be obtained by any processing of a measured unreconstructed power spectrum alone.  However, we are not aware of any way to write this in the same transparent and explicit form as for Eulerian reconstructions.  

Second, our Eulerian growth-shift reconstruction algorithm can be derived from the nonperturbative continuity equation that follows from mass conservation.  It is obtained by expressing the time derivative of the nonlinear mass density in terms of the nonlinear mass density and the linearized velocity density.  This is fully nonperturbative in the mass density, which is important because reconstruction is most interesting in the nonlinear regime where perturbation theory breaks down.  We connect this Eulerian algorithm to the standard Lagrangian reconstruction algorithm by showing that these two algorithms are equivalent in second order perturbation theory.  This theoretical connection can be regarded as a new argument for the robustness and success of the standard Lagrangian BAO reconstruction in the nonlinear regime; it automatically includes 3- and 4-point information in a specific form imposed by the nonlinear continuity equation.  Note however that this point of view is limited by the fact that the nonlinear smoothing of the BAO is largely caused by third order and higher order perturbations, while we only show the equivalence of the algorithms up to second order.  Still, small but important BAO shifts are caused by second order perturbations, and these shifts are reversed in a very similar fashion for Eulerian and Lagrangian reconstructions due to their equivalence at second order.

Third, the methodology of Lagrangian reconstructions may be criticized for intermixing observed data with model assumptions that should actually be tested by the data.  To see this, note that for Lagrangian reconstructions, the observed data is processed by changing observed galaxy positions according to a displacement field that is computed from the same observed galaxy positions under certain model assumptions (including assumptions on halo bias, redshift space distortions, or general relativity).  Thus, the data and assumed fiducial model are mixed together before performing the likelihood analysis that compares the model-dependently-processed data against other models.  Additionally, if the parameters for the displacement field depend on external datasets (e.g.~the CMB to obtain the logarithmic growth rate through $f\sim\Omega_\mathrm{m}^{0.55}$ which holds in general relativity), then the reconstructed density is strictly speaking not independent from this external dataset any more, which may raise concerns about combining constraints on cosmological parameters from these datasets without potentially double-counting information. 

In practice, all of these concerns can of course be modeled or simulated by displacing objects with fiducial models that differ from the true simulated catalogs (e.g.~different values of bias and logarithmic growth rate). The standard reconstruction algorithm seems to be rather robust to such changes.  For example, reference \cite{Padmanabhan2012BAORec} found that using a $20\%$-high linear halo bias for the displacement field shifts the BAO scale by only $1\sigma$ or $2\%$ for their experimental specifications.  However, such shifts may be larger for other survey geometries and need to be checked for every new survey geometry \cite{Padmanabhan2012BAORec}.  Such tests should thus be an integral part of every reconstruction analysis, but running them for every survey geometry and BAO analysis renders the use of reconstruction somewhat cumbersome and computationally expensive.  The Eulerian reconstructions proposed here can ameliorate these complications, because they only depend on the density field of the untransformed observed galaxy positions, and model-independent transformations of this field (like squaring it).   Fiducial bias parameters simply enter the final reconstructed power spectrum by rescaling model-independently measured cross-spectra, so varying these parameters is computationally trivial.

Finally, some of our Eulerian reconstruction algorithms could in principle increase BAO information further, because highly nonlinear scales are shifted differently than in standard reconstruction.  In 2LPT, all reconstruction algorithms considered in this paper are equally well motivated, so it is not clear a priori clear which one recovers most BAO information.  Comparing the performance of the algorithms in simulations is one of the main goals of this paper.

The paper is organized as follows.  We start by deriving two new Eulerian reconstruction schemes that work at the field level:  First, the nonlinear continuity equation is used to derive the Eulerian growth-shift reconstruction algorithm.  Second, we construct an Eulerian $F_2$ reconstruction algorithm that subtracts the second order $F_2$ part of the density.  We then put this into context by modeling the standard Lagrangian BAO reconstruction and some possible alternative Lagrangian reconstruction methods with 2LPT.  Section~\ref{se:RecOverview} provides a high-level overview of the six reconstruction algorithms introduced in the previous sections.  We then proceed with simulations and numerical results, discussing the origin of additional BAO information and the performance of all algorithms.  Finally, we summarize and conclude.  Appendices present Lagrangian modeling calculations, comparisons of reconstructions with 2D slices, a more formal motivation of the Eulerian $F_2$ reconstruction using a Newton-Raphson iteration, and details of the numerical analysis.

\subsection*{Notation and conventions}
We use the following Fourier convention\footnote{The sign in the
exponent is the same as in \cite{padmanabhan0812} but opposite from \cite{marcel1411}.}:
\begin{equation}
  \label{eq:19}
  f(\vk) = \int\d^3\vx\; e^{-i\vk\cdot\vx}f(\vx),\qquad
  f(\vx) = \int\frac{\d^3\vk}{(2\pi)^3}\; e^{i\vk\cdot\vx}f(\vk),
\end{equation}
i.e~the Fourier transform of the gradient of a function is
\begin{equation}
  \label{eq:20}
  [\nabla f](\vk) = i\vk f(\vk).
\end{equation}
Throughout the paper, $\vq$ stands for Lagrangian configuration-space coordinates and $\vx$
for Eulerian configuration-space coordinates, while $\vk$ and $\vk_i$ denote Fourier-space
wavevectors. 
Coordinates and wavenumbers are comoving.  
To shorten expressions we write
\begin{equation}
  \label{eq:DefShortInt}
  \int_{\vk_{i}}^* \equiv \int\frac{\d^3\vk_1}{(2\pi)^3}\,\frac{\d^3\vk_2}{(2\pi)^3}\,
(2\pi)^3\delta_D(\vk-\vk_1-\vk_2).
\end{equation}
Throughout the paper, second order Lagrangian Perturbation Theory or 2LPT refers to expanding all terms in LPT  consistently in the linear density $\delta_0$ up to $\mathcal{O}(\delta_0^2)$, i.e.~we do not include corrections beyond second order that would follow e.g.~by using the cumulant expansion theorem (which would be an interesting extension).

\section{Eulerian Growth-Shift Reconstruction from Nonlinear
Continuity Equation}
\label{se:EGS}
\subsection{Dark Matter}
We propose to reverse the nonlinear time evolution of structures by starting with the
observable nonlinear density $\delta(\vx,\eta)$ today and going back
in time by some  time interval $\Delta\eta$ using a Taylor expansion in time, 
\begin{equation}
  \label{eq:delta_rec_def}
\delta(\vx,\eta-\Delta\eta)
\approx \delta(\vx,\eta) -
  \Delta\eta\, \partial_\eta\delta(\vx,\eta) \equiv \delta_\mathrm{rec}(\vx),
\end{equation}
where  $\eta$ is conformal
time, $\eta=\int_0^a\tfrac{\d\ln a'}{a'H(a')}$. The time derivative of
the (nonlinear) density can be obtained from the nonperturbative
continuity equation that follows from mass conservation.  It states that any change of the mass density in a particular volume element with time must be caused by an equivalent in- or out-flow of mass (described by the divergence of the mass-weighted velocity or momentum),
\begin{equation}
  \label{eq:61}
  \partial_\eta\delta + \nabla\cdot\left[
(1+\delta)\vv \right] = 0.
\end{equation}
Here,  $\vv$ is the (nonlinear) peculiar velocity perturbation
which we assume to be curl-free with $\theta\equiv \nabla\cdot\vv$,
i.e.~$\vv(\vk)=-i\vk\theta(\vk)/k^2$. Writing out the spatial
derivative gives
\begin{equation}
  \label{eq:delta_prime}
  \partial_\eta\delta(\vx,\eta) = -\nabla\cdot\vv(\vx,\eta) -
  \vv(\vx,\eta)\cdot\nabla\delta(\vx,\eta) 
- \delta(\vx,\eta)\nabla\cdot\vv(\vx,\eta).
\end{equation}
If we know the density and velocity fields then we can directly use this in
\eqq{delta_rec_def} to reconstruct the density at an earlier time.

In practice, the locations of objects are easier to observe than
their peculiar velocities, so that the mass density is typically much better known
than the velocity field.  At linear order we can infer the
velocity field from the mass density,
\begin{equation}
  \label{eq:v0_intermsof_delta0}
  \vv_0(\vk,\eta) = f\mathcal{H}\frac{i\vk}{k^2}\delta_0(\vk,\eta).
\end{equation}
This relation is obtained from the perturbative expansion
\begin{equation}
  \label{eq:57}
\delta(\vx,\eta)=D(\eta)\delta_0(\vx)+\mathcal{O}(D^2\delta_0^2),
\end{equation}
where time and spatial dependence are described by the linear growth function $D(\eta)$ and the linear perturbation $\delta_0(\vx)$, respectively.  The linearized continuity equation then becomes
\begin{equation}
  \label{eq:div_v0}
  \nabla\cdot\vv_0(\vx,\eta) = -\delta_0(\vx)\partial_\eta D(\eta) = -f\mathcal{H}\delta_0(\vx,\eta),
\end{equation}
which implies \eqq{v0_intermsof_delta0}.  Here, $f=\d\ln D/\d\ln a$ is the logarithmic growth rate and
$\mathcal{H}=aH$ is the comoving Hubble parameter.  The linear density $\delta_0$ in \eqq{v0_intermsof_delta0} is unknown in observations, but it can be approximated by the observed nonlinear density if small scales corrupted by nonlinearities are smoothed out:
\begin{equation}
  \label{eq:62}
\delta_0(\vk)\approx  \delta_R(\vk) \equiv W_R(k)\delta(\vk),
\end{equation}
where $W_R$ is a smoothing kernel that is unity on large scales $k\lesssim R^{-1}$ and vanishes on small scales $k\gtrsim R^{-1}$. The smoothing scale $R$ should be chosen close to the scale where linear perturbation theory breaks down.  The corresponding approximate velocity is
\begin{equation}
  \label{eq:67}
  \vv(\vx,\eta) \approx -f\mathcal{H} \vs(\vk,\eta),
\end{equation}
where we defined
\begin{equation}
  \label{eq:vs_def2}
  \vs(\vk,\eta) \equiv - \frac{i\vk}{k^2}W_R(k)\delta(\vk,\eta),
\end{equation}
which coincides with the negative Zeldovich displacement.
The velocity divergence is simply
\begin{equation}
  \label{eq:68}
  \nabla\cdot\vv(\vx,\eta)\approx -f\mathcal{H}\, \delta_R(\vx,\eta).
\end{equation}
The density time derivative \eq{delta_prime} then becomes
\begin{equation}
  \label{eq:delta_prime_intermsof_deltaNL}
  \partial_\eta \delta(\vx,\eta) = f\mathcal{H}\Big[
\delta_R(\vx,\eta) +\vs(\vx,\eta)\cdot\nabla\delta(\vx,\eta) + \delta(\vx,\eta)\delta_R(\vx,\eta)
\Big],
\end{equation}
so that the Euler reconstructed field of \eqq{delta_rec_def} is given by
\begin{equation}
  \label{eq:delta_rec_with_DeltaEta}
  \delta_\mathrm{rec}(\vx) = \delta(\vx,\eta) - \Delta\eta\,
f\mathcal{H}\,\Big[
\vs(\vx,\eta)\cdot\nabla\delta(\vx,\eta) + \delta(\vx,\eta)\delta_R(\vx,\eta)
\Big].
\end{equation}
We dropped the first term in the square brackets of \eqq{delta_prime_intermsof_deltaNL} because it describes linear time evolution but we only want to reverse nonlinear time evolution when reconstructing the linear density.  The go-back time $\Delta\eta$ is a free parameter, but from now on
we choose $\Delta\eta=(f\mathcal{H})^{-1}$ so that the coefficient of
the square brackets in \eqq{delta_rec_with_DeltaEta} becomes $-1$.
We call the resulting algorithm Eulerian growth-shift reconstruction, or short `EGS',
\begin{equation}
  \label{eq:delta_EGS_def}
  \delta_\mathrm{EGS}^\rec(\vx) =  \delta(\vx,\eta) -
\underbrace{\vs(\vx,\eta)\cdot\nabla\delta(\vx,\eta)}_\mathrm{shift} 
- \underbrace{\delta(\vx,\eta)\delta_R(\vx,\eta)}_\mathrm{growth},
\end{equation}
because it subtracts nonlinear growth and shift from the nonlinear density at the field level.
As desired, the right hand side of \eqq{delta_EGS_def} depends only on the mass density $\delta$ and not the
velocity so it can be easily obtained from observations.

The shift and growth term in \eqq{delta_EGS_def} are
quadratic in the nonlinear unreconstructed density.  The 2-point correlation function
or power spectrum of the
reconstructed density thus involves specific 2-, 3- and 4-point statistics of the
unreconstructed density,
  \begin{equation}
    \label{eq:P_EGS}
\hat P_{\delta_\mathrm{EGS}^\rec,\delta_\mathrm{EGS}^\rec}(k) =
\underbrace{\hat P_{\delta,\delta}(k)}_\mathrm{2pt}
\underbrace{-
2\hat P_{\delta^2 , \delta} (k)
-2\hat P_{ \vs\cdot\nabla\delta , \delta}(k)
}_\mathrm{3pt}
+
\underbrace{\hat P_{ \delta^2 ,\delta^2 }(k)
+\hat P_{ \vs\cdot\nabla\delta , \vs\cdot\nabla\delta}(k)
+2\hat P_{ \vs\cdot\nabla\delta ,\delta^2}(k)}_\mathrm{4pt},
  \end{equation}
where we sloppily wrote $\delta^2$ for $\delta_R(\vx)\delta(\vx)$.   This makes it very transparent how reconstruction adds information to the power spectrum by combining information from 2-, 3- and 4-point statistics of the unreconstructed density.

 In fact, the 3-point cross-spectra of $\delta^2$ or $\vs\cdot\nabla\delta$ with $\delta$ in \eqq{P_EGS} do not just probe the 3-point function, but they represent nearly-optimal maximum-likelihood estimators for the amplitudes of the contributions to the tree-level 3-point function due to nonlinear gravitational growth and shift \cite{marcel1411}.  Roughly, they measure the projection of the observed 3-point function on tree-level expectations, and are thus expected to capture a significant fraction of the full 3-point information.

Note that the standard Lagrangian reconstruction method also employs the continuity equation.  However, it is only used there in its fully linearized form, where both mass density and velocity are linearized, to obtain the first order velocity or displacement $\vs_0=-i\vk/k^2\delta_0$.  We also use this to approximate the velocity.  In contrast to previous literature, we then plug this linearized velocity back into the nonlinear continuity  equation without linearizing the mass density.  The resulting density time derivative  \eq{delta_prime_intermsof_deltaNL} is \emph{nonperturbatively} correct in the unsmoothed nonlinear density (appearing on the left hand side and in the last two summands on the right hand side of \eqq{delta_prime_intermsof_deltaNL}). It only uses perturbation theory to approximate the factors in \eqq{delta_prime} that involve the velocity, leading to $\vs$  and $\delta_R$ in Eqs.~\eq{delta_prime_intermsof_deltaNL} and \eq{delta_EGS_def}.   Thus, the crucial difference to previous approaches is that we use a form of the continuity equation that is linearized only in the velocity but fully nonperturbative in the mass density.  Since reconstruction is most useful on small scales where perturbation theory breaks down, this nonperturbative motivation is very valuable.

An extension that goes beyond the scope of this paper would be to use better approximations for the velocity e.g.~by using forward-modeling techniques or iterative estimators.

\subsection{Halos}

Our derivation easily extends to biased tracers like halos if we assume that they
define a density field that exists even at early times and that halos (or density peaks)
are conserved over time.\footnote{\changed{We do not discuss the validity of this assumption, but see e.g.~\cite{Chan1201,tobias1405} for related recent discussions.  We also leave the extension to nonlinear and nonlocal halo bias to future work, noting that we focus on dark matter in this paper.}}  The continuity equation for the halo density is then
\begin{equation}
  \label{eq:59}
  \partial_\eta\delta_h + \nabla\cdot\left[(1+\delta_h)\mathbf{v}_h\right]=0,
\end{equation}
so that
\begin{equation}
  \label{eq:52}
  \partial_\eta\delta_h = -\nabla\cdot\mathbf{v}_h -
  \mathbf{v}_h\cdot\nabla\delta_h 
 - \delta_h\nabla\cdot\mathbf{v}_h.
\end{equation}
We now proceed to approximate the halo velocity. Assuming that the
halo velocity follows the dark matter (DM) velocity with some non-stochastic velocity bias, $\vv_h=b_v\vv$,
and approximating the DM velocity as above using the linearized
continuity equation, gives
\begin{equation}
  \label{eq:58}
  \vv_h(\vk) = b_v\vv(\vk) =f\mathcal{H} b_v \frac{i\vk}{k^2}W_R(k)\delta(\vk).
\end{equation}
In practice we only know the halo density $\delta_h$ and not the DM
density $\delta$, but ignoring higher order bias the two can be
related by $\delta_h=b_1\delta$ so that
\begin{equation}
  \label{eq:60}
  \vv_h(\vk) = f\mathcal{H}\frac{b_v}{b_1} \frac{i\vk}{k^2}W_R(k)\delta_h(\vk).
\end{equation}
The reconstruction algorithm is thus the same as for the DM case, with an additional  prefactor $b_v/b_1$, which rescales the halo density to the halo velocity, 
\begin{equation}
  \label{eq:EulerRec_DeltaEta_halos}
  \delta_{h,\mathrm{EGS}}^\mathrm{rec}(\vx) = \delta_h(\vx,\eta) - \Delta\eta\,
f\mathcal{H}\,\frac{b_v}{b_1}\,\Big[
\vs_h(\vx,\eta)\cdot\nabla\delta_h(\vx,\eta) + \delta_h(\vx,\eta)\delta_{h,R}(\vx,\eta)
\Big],
\end{equation}
where we dropped the term that reverses linear time evolution as in \eqq{delta_rec_with_DeltaEta}.

The power spectrum of the reconstructed halo density \eq{EulerRec_DeltaEta_halos} is a combination of halo 2-, 3- and 4-point statistics, similar to the DM case of \eqq{P_EGS}.  The fiducial bias parameters used for the reconstruction enter simply as coefficients of halo cross-spectra.  Changing these fiducial bias parameters would simply up- or down-weight the corresponding cross-spectra, making it computationally trivial to obtain the reconstructed power spectrum for varying fiducial bias parameters if all cross-spectra are pre-computed once.  Such changes are more cumbersome for Lagrangian reconstructions, where for every new fiducial value of a bias parameter all particles need to be displaced by the corresponding updated displacement field.

\section{Eulerian F2 Reconstruction to Remove
    2nd Order Perturbations}
\label{se:EF2Rec}

While the previous section was based on linearizing only the velocity but not the mass density in the continuity equation, we now proceed by linearizing both the mass density and velocity in all equations of motion for the DM fluid (i.e.~the continuity, Poisson and Euler equation).  Working at second order, we construct a different Eulerian reconstruction algorithm that aims to reverse the full second order part of the nonlinear density.  As shown in Appendix~\ref{se:NewtonRaphson}, the same algorithm can be derived slightly more formally by applying the Newton-Raphson method to find the linear density that generates a given observed density at second order.

\subsection{Dark Matter}
Solving the equations of motion for DM perturbatively in the density gives the solution $\delta=\delta_0+\delta^{(2)}+\mathcal{O}(\delta_0^3)$ where $\delta_0$ is the linear density and the second order part is given as a sum of growth, shift and tidal term (e.g.~\cite{Bouchet1992ApJ...394L...5B,tobias1201,SherwinZaldarriaga2012,marcel1411}):
\begin{equation}
  \label{eq:delta2_real_space}
\delta^{(2)}(\vx) = \underbrace{\frac{17}{21} \delta_0^2(\vx)}_\mathrm{growth} -
\underbrace{\vPsi_0(\vx)\cdot\nabla\delta_0(\vx)}_\mathrm{shift} + \underbrace{\frac{4}{21}K_0^2(\vx)}_\mathrm{tidal}.
\end{equation}
The linear displacement field is\footnote{The Fourier convention used here differs from
  e.g.~\cite{marcel1411} so that
  $\vPsi_\mathrm{here}=-\vPsi_\mathrm{there}$.  The convention used
  here agrees with \cite{padmanabhan0812}.}
\begin{equation}
  \label{eq:vPsi0}
  \vPsi_0(\vk) = \frac{i\vk}{k^2}\delta_0(\vk)
\end{equation}
and the tidal term is given by contracting the tidal tensor
\begin{equation}
  \label{eq:Kij_lin}
  K_0^{ij}(\vk) =  \left(\frac{k_ik_j}{k^2} - \frac{1}{3}\delta_{ij}^{\mathrm{(K)}}
\right)\delta_0(\vk)
\end{equation}
to
\begin{equation}
  \label{eq:tidal_x_def}
  K^2_0(\vx) = \frac{3}{2}K_0^{ij}(\vx)K_0^{ij}(\vx),
 \end{equation}
where $\delta_{ij}^{\mathrm{(K)}}$ is the Kronecker delta. In Fourier space this becomes
\begin{equation}
  \label{eq:56}
  K^2_0(\vk) = \int_{\vk_i}^*
  \mathsf{P}_2(\hat{\vk}_1\cdot \hat{\vk}_2)\delta_0(\vk_1)\delta_0(\vk_2),
\end{equation}
where $\mathsf{P}_2$ denotes the $l=2$ Legendre polynomial,
\begin{equation}
  \label{eq:35}
  \mathsf{P}_2(\mu) = \frac{3}{2}\left(\mu^2 -\frac{1}{3}\right).
\end{equation}
The second-order density of \eqq{delta2_real_space} is often written as a convolution in Fourier space,  
\begin{equation}
  \label{eq:delta_2_k}
  \delta^{(2)}(\vk) 
= \int \frac{\d^3 k_1}{(2\pi)^3} F_2(\vk_1, \vk-\vk_1)\delta_0(\vk_1)\delta_0(\vk-\vk_1)
= \int_{\vk_i}^* F_2(\vk_1,\vk_2)\delta_0(\vk_1)\delta_0(\vk_2),
\end{equation}
where the shorthand notation $\int_{\vk_i}^*$ is defined by \eqq{DefShortInt} and the symmetrized $F_2$ kernel contains growth, shift and tidal parts analogously to \eqq{delta2_real_space},
\begin{equation}
  \label{eq:66}
F_2(\vk_1,\vk_2) =  \frac{17}{21}
+ \frac{1}{2}\,\left(\frac{k_1}{k_2}+\frac{k_2}{k_1}\right)
\hat{\vk}_1\cdot\hat{\vk}_2 +
\frac{4}{21}\mathsf{P}_2(\hat{\vk}_1\cdot\hat{\vk}_2).
\end{equation}

To restore linear information from the nonlinear observed density we
try to estimate the second order part \eq{delta2_real_space} and
subtract it out.\footnote{More formally, one can model the nonlinear
  density in terms of the linear one as $\delta=\delta_0+F_2[\delta_0,\delta_0]+\cdots$. Then the goal is to get $\delta_0$ from
  $\delta$. This can be done with the Newton-Raphson root-finding method, see Appendix~\ref{se:NewtonRaphson}. The second step of
  this corresponds to subtracting $F_2[\delta,\delta]$ from
  $\delta$, which agrees with the intuitive picture of subtracting out
  nonlinearities from the nonlinear density.}  In practice we do not
know the linear density $\delta_0$ that enters
\eqq{delta2_real_space}, but we can approximate it by the smoothed 
nonlinear density because linear and nonlinear densities agree on large
scales.  There is some freedom which fields in \eqq{delta2_real_space}
should be smoothed or linearized.  Motivated by the asymmetric
smoothing that followed from the continuity equation in the last
section, we choose to smooth one of the two fields entering the growth
term $\delta^2$ and the tidal term $K^2$, and we only smooth $\vPsi$
but not $\nabla\delta$ in the shift term.  This defines the Eulerian $F_2$ reconstruction
`EF2':
\begin{equation}
  \label{eq:delta_EF2_def}
  \delta_\mathrm{EF2}^\rec(\vx) \equiv \delta(\vx) -\underbrace{\frac{17}{21}\delta_R(\vx)\delta(\vx)}_\mathrm{growth}
  -\underbrace{\vs(\vx)\cdot\nabla\delta(\vx)}_\mathrm{shift} -\underbrace{\frac{4}{21}K^2_R(\vx)}_\mathrm{tidal},
\end{equation}
where $\delta_R$ is the smoothed nonlinear density, $\vs$ is the
negative smoothed Zeldovich displacement given by \eqq{vs_def2}, and
$K^2_R(\vx)\equiv\tfrac{3}{2}K^{ij}K^{ij}_R$ is a partially smoothed
tidal term where $K^{ij}$ is defined as in \eqq{tidal_x_def} but using
the nonlinear density $\delta$ instead of the linear density
$\delta_0$, and $K_R^{ij}$ is defined in terms of the smoothed
nonlinear density $\delta_R$ instead of $\delta_0$.

The Eulerian $F_2$ reconstruction of \eqq{delta_EF2_def} is similar to the
growth-shift reconstruction of \eqq{delta_EGS_def}; the shift terms
$\vs\cdot\nabla\delta$ are equal in both methods, while the growth term
$\delta_R\delta$ has a slightly different coefficient, $17/21$
instead of $1$, and the $F_2$
reconstruction subtracts an additional tidal part. Both methods agree
with the unreconstructed density on very large scales because the
combination of quadratic fields that is subtracted vanishes there
(this follows by considering the squeezed limit of the corresponding
kernels in Fourier space).

\subsection{Halos}
It is straightforward to extend this algorithm to halos if a specific bias relation between DM and halos is assumed. For example, if only linear and second order bias are taken into account (neglecting tidal and velocity bias), we have
\begin{equation}
  \label{eq:79}
  \delta_h(\vx) = b_1\big[\delta_0(\vx) + F_2[\delta_0,\delta_0](\vx)\big] + b_2\left[\delta_0^2(\vx)-\la\delta_0^2(\vx)\ra\right]+\mathcal{O}(\delta_0^3),
\end{equation}
where we wrote $F_2[\delta_0,\delta_0](\vx)\equiv \delta^{(2)}(\vx)$ for the Fourier transform of \eqq{delta_2_k}.
At linear order $\delta_0=\delta_h/b_1$, so that subtracting second-order parts from the nonlinear halo density gives (ignoring smoothing, which should be included)
\begin{equation}
  \label{eq:80}
  \delta^\mathrm{rec}_{h,\mathrm{EF2}}(\vx) = 
\delta_h(\vx) - \frac{1}{b_1}F_2[\delta_h,\delta_h](\vx) - \frac{b_2}{b_1^2}\left[\delta_h^2(\vx)-\la\delta_h^2(\vx)\ra\right].
\end{equation}
This is an estimator for the linear part $b_1\delta_0$ of the halo density given the nonlinear halo density $\delta_h$.  Note that this is similar to the EGS reconstruction for halos given by \eqq{EulerRec_DeltaEta_halos}, but has slightly different coefficients and an additional tidal term coming from the $F_2$ kernel.

The algorithm can be extended to redshift space by subtracting the second order part of the redshift space density, which can be obtained by modifying the real space $F_2$ kernel to a redshift space version $Z_2$ (see~\cite{BernardeauReview} for a review).  The form of the resulting terms is similar to that of real space growth, shift and tidal terms from the $F_2$ kernel, but derivative operations differ in detail because of redshift space distortions. Similarly, the EGS algorithm of the previous section could be extended to redshift space by inverting the redshift space continuity equation.  However, we leave these extensions for future work and restrict ourselves to reconstructions for DM in real space in this paper, which is somewhat cleaner and simpler.  It should be noted that in contrast to our Eulerian algorithms, the standard Lagrangian BAO reconstruction algorithm, introduced almost 10 years ago \cite{EisensteinRecSims0604362}, has been extended and applied to redshift space, e.g.~in reference \cite{Padmanabhan2012BAORec}.

\section{Lagrangian Reconstructions}
\label{se:LagrangianRecs}
\subsection{Possibilities}
\label{se:possible_lagrangian_recs}

The standard BAO reconstruction algorithm, proposed by \cite{EisensteinRecSims0604362} and denoted Lagrangian growth-shift `LGS' reconstruction in this paper, works as follows.  First, the particles of a clustered catalog are displaced by the negative Zeldovich displacement $\vs$ to obtain a `displaced' catalog, whose density is denoted $\delta_d[\vs]$.  Similarly, particles of a random catalog are shifted by the same displacement field to give a 'shifted` catalog with density $\delta_s[\vs]$.  Then the density difference $\delta_d[\vs]-\delta_s[\vs]$ gives the reconstructed density and suppresses only the nonlinear but not the linear part of the density (see Appendix~\ref{se:LagrangianTheory}).  

As a simple extension we also consider clustered and random catalogs shifted by the \emph{positive} Zeldovich displacement, $-\vs$.  We denote the resulting densities $\delta_d[-\vs]$ and $\delta_s[-\vs]$.     At first order in $\delta_0$ we have (following from Eqs.~\eq{delta_k_from_Psi_k}-\eq{delta_s_shifted} in Appendix~\ref{se:LagrangianTheory} and $i\vk\cdot\vs(\vk)=W_R(k)\delta_0(\vk)$)
\begin{eqnarray}
  \label{eq:64}
  \delta_d^{(1)}[\vs](\vk) &=& [1-W_R(k)]\delta_0(\vk) \\
  \delta_s^{(1)}[\vs](\vk) &=& -W_R(k)\delta_0(\vk) \\
  \delta_d^{(1)}[-\vs](\vk) &=& [1+W_R(k)]\delta_0(\vk) \\
  \delta_s^{(1)}[-\vs](\vk) &=& W_R(k)\delta_0(\vk). 
\end{eqnarray}
There are several possible combinations of these four shifted/displaced catalogs that change the nonlinear part of the density while keeping the linear part unchanged (for arbitrary smoothing):
\begin{eqnarray}
  \label{eq:rec_combi1}
\delta_\mathrm{LGS}^\mathrm{rec} \equiv \delta_d[\vs]-\delta_s[\vs]   &=&\delta_0 + \mathcal{O}(\delta_0^2)\\
 \tfrac{1}{2}\left\{\delta_d[\vs]+\delta_d[-\vs]\right\}   &=&\delta_0 + \mathcal{O}(\delta_0^2)\\
 \delta_d[\vs]+\delta_s[-\vs]   &=&\delta_0 + \mathcal{O}(\delta_0^2)\\
 \delta_s[\vs]+\delta_d[-\vs]   &=&\delta_0 +
 \mathcal{O}(\delta_0^2)\\
\label{eq:rec_combi5}
\delta_\mathrm{LRR}^\rec\equiv\delta - c\left\{ \delta_s[\vs]+\delta_s[-\vs]\right\}   &=&\delta_0 +\mathcal{O}(\delta_0^2)\\
  \label{eq:rec_combi6}
 \delta_d[-\vs]-\delta_s[-\vs]   &=&\delta_0 + \mathcal{O}(\delta_0^2),
\end{eqnarray}
where $c$ in \eqq{rec_combi5} is an arbitrary constant.  Beyond first order these
combinations generally differ from each other.  

For reconstructions of the linear density, the nonlinear contributions to the nonlinear density should be
suppressed, particularly the growth and shift part.  Modeling all combinations by 2LPT shows that this is only the case for two combinations (see Appendix~\ref{se:alt_rec_2nd_order}): First, the LGS reconstruction algorithm of \eqq{rec_combi1}, which is just the standard BAO reconstruction algorithm.  Second, a new Lagrangian \emph{random-random} `LRR' reconstruction algorithm, which is given by \eqq{rec_combi5} if we choose $c=1/2$ (see Appendix~\ref{se:alt_rec_2nd_order}),
\begin{equation}
  \label{eq:LRRdef}
  \delta^\mathrm{rec}_\mathrm{LRR} = \delta - \frac{1}{2}\left\{\delta_s[\vs]+\delta_s[-\vs]\right\}.
\end{equation}
In this case, the reconstructed density is obtained as follows.  A random catalog is first shifted by the negative Zeldovich displacement $\vs$ to get $\delta_s[\vs]$.  The original random catalog is also shifted by the positive Zeldovich displacement $-\vs$ to get $\delta_s[-\vs]$.  Then the mass densities of these two shifted random catalogs are computed and added together.  Finally, half of that sum is subtracted from the original density of the clustered catalog to obtain the reconstructed density.

\subsection{Lagrangian Growth-Shift and Random-Random Reconstructions}
\label{se:LGSandLRR}
We now discuss the LGS and LRR algorithms in more detail.
The theoretical 2LPT expression for the standard LGS
method is (see \eqq{delta_LGS_2nd_order_appendix}, using superscript `$\theo$' to denote theoretical densities)
\begin{equation}
  \label{eq:delta_LGS_2nd_order}
  \delta_\mathrm{LGS}^{\rec,\theo}(\vk) = \delta^\theo(\vk) - 
\int_{\vk_i}^*
\left[1+\frac{k_1}{k_2}\hat{\vk}_1\cdot\hat{\vk}_2\right]
W_R(k_2)\delta_0(\vk_1)\delta_0(\vk_2),
\end{equation}
or more succinctly in configuration space,
\begin{equation}
  \label{eq:delta_LGS_x_LPTfinal}
  \delta_\mathrm{LGS}^{\rec,\theo}(\vx) = \delta^\theo(\vx) - \delta_{0,R}(\vx)\delta_0(\vx) -
  \vs_0(\vx)\cdot\nabla \delta_0(\vx).
\end{equation}
Here, $\delta_{0,R}$ is the smoothed linear density and $\vs_0$ is the
negative smoothed Zeldovich displacement computed from the linear
density $\delta_0$ analogously to \eqq{vs_def2}.  Thus, in this 2LPT
model, LGS reconstruction subtracts growth and shift terms from the
nonlinear density $\delta$, so that in 2LPT theory the reconstructed density of the Lagrangian
growth-shift method of \eqq{rec_combi1} agrees with that of the Eulerian
growth-shift method of \eqq{delta_EGS_def}, derived from the
continuity equation for the nonlinear mass density.    

Note that the statement here is that the reconstructed \emph{densities} agree at second order LPT.  For the reconstructed \emph{power spectrum} this means that the tree level and `22' loop contributions agree between the methods, whereas the `13' loop correction (which is of the same order as the `22' contribution) can differ between these algorithms because the reconstructed densities may differ at third order, which we do not model here; see Section~\ref{se:13_22_split} for further discussion of this point using simulations.

Details of the 2LPT calculation that leads to Eqs.~\eq{delta_LGS_2nd_order} and \eq{delta_LGS_x_LPTfinal} are given in Appendix~\ref{se:LagrangianTheory}.  These results only hold after introducing a new correction term that arises from modeling the displaced clustered catalog such that the displacement field $\vs$ is evaluated at Eulerian  instead of Lagrangian positions; see Appendices~\ref{se:LagrangianTheory} and \ref{se:SliceComparisons}.

Similarly, modeling the random-random reconstruction of \eqq{LRRdef} with
2LPT gives
(see Appendix \ref{se:alt_rec_2nd_order})
\begin{equation}
\label{eq:delta_LRR_2nd_order}
  \delta_\mathrm{LRR}^{\mathrm{rec},\theo}(\vk) =
\delta^\theo(\vk)- \int_{\vk_i}^* 
\left[\frac{2}{3}
+ \frac{1}{2}\,\left(\frac{k_1}{k_2}+\frac{k_2}{k_1}\right)
\hat{\vk}_1\cdot\hat{\vk}_2 +
\frac{1}{3}\mathsf{P}_2(\hat{\vk}_1\cdot\hat{\vk}_2)\right]
W_R(k_1)W_R(k_2)
\delta_0(\vk_1)\delta_0(\vk_2),
\end{equation}
The square brackets correspond to the part of the $F_2$ kernel that is
generated by $[L^{(1)}]^2$ in 2LPT; see \eqq{F2_11_LPT}.  In configuration space this becomes
\begin{equation}
  \label{eq:delta_LRR_x_LPTfinalMainText}
    \delta_\mathrm{LRR}^{\rec,\theo}(\vx) =\delta^\theo(\vx) -
    \frac{2}{3}\delta^2_{0,R}(\vx) -
    \vs_0(\vx)\cdot\nabla\delta_{0,R}(\vx) 
-\frac{1}{3}K_{0,RR}^2(\vx),
\end{equation}
where $K_{0,RR}^2(\vx)=\tfrac{3}{2}K^{ij}_{0,R}K^{ij}_{0,R}$ is
defined in terms of the smoothed linear density.

The nonlinear density before reconstruction can be modeled by 
\begin{equation}
  \label{eq:77}
  \delta^\theo(\vx) = \delta_0(\vx)+\frac{17}{21}\delta_0^2(\vx)
-\vPsi_0(\vx)\cdot\nabla\delta_0(\vx)+\frac{4}{21}K^2_{0}(\vx)
\end{equation}
at second order LPT or SPT.  On large scales, $\delta_0\approx
\delta_{0,R}\approx\delta_R$ and $\vPsi_0\approx -\vs_0\approx -\vs$.
Thus, both reconstruction methods LGS and LRR cancel the shift term
$\vPsi\cdot\nabla\delta$ exactly on large scales as desired. 

However, even at second order the reconstructions do not perfectly
recover the linear density $\delta_0$.
For the LGS method,  the residual second order part of the
density after reconstruction is
\begin{align}
\label{eq:delta_LGS_2nd_order_residual}
  \delta_\mathrm{LGS}^{\rec,\theo}(\vk) - \delta_0(\vk) =
\int_{\vk_i}^* 
\Bigg\{
\left[\frac{17}{21}-W_R(k_2)\right]
+\left[1-W_R(k_2)\right]
\frac{k_1}{k_2}
\hat{\vk}_1\cdot\hat{\vk}_2
+\frac{4}{21}\mathsf{P}_2(\hat{\vk}_1\cdot\hat{\vk}_2)
\Bigg\}\delta_0(\vk_1)\delta_0(\vk_2).
\end{align}
On large scales, where $W_R(k_1)\rightarrow 1$, reconstruction reduces the growth part from $17/21$ to $-4/21$ and it fully removes the shift term, but it does not change the tidal
part of the nonlinear density.  On smaller scales, where
$W_R(k_1)\rightarrow 0$, reconstruction does not change the second order part of the density, which is reasonable because the linear displacement field cannot be reliably estimated from the nonlinear density any more on these scales, and still attempting to do so may inadvertently enhance nonlinearities.\footnote{Reversely, this fact could be exploited by applying reconstruction only to modes or local environments for which the velocity or displacement can be estimated reliably \cite{Achitouv2015}.}

For the LRR method the residual second order density is similar,
\begin{align}
\nonumber
  \delta_\mathrm{LRR}^{\rec,\theo}(\vk) -\delta_0(\vk) = \int_{\vk_i}^*
\Bigg\{&
\left[\frac{17}{21}-\frac{14}{21}W_R(k_1)W_R(k_2)\right]
+\left[1-W_R(k_1)W_R(k_2)\right]
\frac{k_1}{k_2}
\hat{\vk}_1\cdot\hat{\vk}_2\\
\label{eq:delta_LRR_2nd_order_residual}
&\,
+\left[\frac{4}{21}-\frac{7}{21}W_R(k_1)W_R(k_2)\right] 
\mathsf{P}_2(\hat{\vk}_1\cdot\hat{\vk}_2) 
\Bigg\}\delta_0(\vk_1)\delta_0(\vk_2).
\end{align}
For large scale modes, the growth term is reduced from $17/21$ to
$3/21$, the shift term is removed entirely, and the tidal term is changed
from $4/21$ to $-3/21$; on small scales the second order part is again
unchanged. Note that the LRR method involves two smoothing kernel
factors that smoothen both displacement and density fields.  This
aggressive smoothing may remove small-scale density modes that could
actually still be used if they were displaced by large-scale
displacements, at least for applications like measuring the BAO scale.

On the very largest scales, $k\rightarrow 0$, the second order residual densities given by the right hand sides of Eqs.~\eq{delta_LGS_2nd_order_residual} and \eq{delta_LRR_2nd_order_residual} vanish.   To see this, note that for $k\ll k_1,k_2$, we have $\vk_2=\vk-\vk_1\approx -\vk_1$ corresponding to a squeezed  triangle configuration, so that $k\ll k_1\approx k_2$. Then we get $\mathsf{P}_2(-1)= 1$.  For $k_1,k_2\lesssim R^{-1}$, $W_R(k_i)=1$, and the right hand sides of Eqs.~\eq{delta_LGS_2nd_order_residual} and \eq{delta_LRR_2nd_order_residual} become $-4/21+4/21=0$ and $3/21-3/21=0$, respectively.  In the regime $k_1,k_2\gtrsim R^{-1}$, $W_R(k_i)=0$, and the right hand sides of Eqs.~\eq{delta_LGS_2nd_order_residual} and \eq{delta_LRR_2nd_order_residual} both become $17/21-1+4/21=0$.

\subsection{Lagrangian F2 Reconstruction}
\label{se:LF2}
The 2LPT theory expressions \eq{delta_LGS_2nd_order} and \eq{delta_LRR_2nd_order} for the LGS and LRR methods (or the residual equations \eq{delta_LGS_2nd_order_residual} and \eq{delta_LRR_2nd_order_residual}) can be used to
form a linear combination of these two reconstruction algorithms that
removes the full $F_2$ kernel (on scales that are not smoothed out) and still leaves the
first order part unchanged,\footnote{Ignoring smoothing, there is
  exactly one linear combination
  $c_1\delta_\mathrm{LGS}^\rec+c_2\delta_\mathrm{LRR}^\rec$ that works. To keep
  the correct shift term we need $c_1+c_2=1$. The growth
  term imposes $c_1+\tfrac{2}{3}c_2=\frac{17}{21}$.  The tidal
  term is fixed by the fact that the kernel needs to vanish in the
  squeezed limit (or it imposes $\frac{1}{3}c_2=\frac{4}{21}$). The only solution is $c_1=\frac{3}{7}$ and $c_2=\frac{4}{7}$.}
\begin{equation}
  \label{eq:delta_LF2}
\delta_\mathrm{LF2}^\rec\equiv\frac{3}{7}\delta_\mathrm{LGS}^\rec
+ \frac{4}{7} \delta_\mathrm{LRR}^\rec
=
\frac{4}{7}\delta + \frac{3}{7}\delta_d[\vs] -
 \frac{5}{7}\delta_s[\vs]-\frac{2}{7}\delta_s[-\vs],
\end{equation}
where `LF2' stands for Lagrangian $F_2$ reconstruction.
Indeed, at second order in LPT
\begin{align}
\nonumber
  \delta_\mathrm{LF2}^{\rec,\theo}(\vk)= \delta^\theo(\vk)-\int_{\vk_i}^* W_R(k_2)&
\bigg\{
\left[\frac{9}{21}+\frac{8}{21}W_R(k_1)\right]
+ 
\left[\frac{3}{7}+\frac{4}{7}W_R(k_1)\right]
\frac{k_1}{k_2}\hat{\vk}_1\cdot\hat{\vk}_2\\
&\,  
\label{eq:delta_LF2_2nd_order}
+
\frac{4}{21}W_R(k_1)\mathsf{P}_2(\hat{\vk}_1\cdot\hat{\vk}_2)
\bigg\}\delta_0(\vk_1)\delta_0(\vk_2),
\end{align}
where the kernel in curly brackets equals $F_2$ for large scales $k_i\lesssim R^{-1}$, where
$W_R(k_i)=1$, i.e.~the full second order $F_2$ part of the
nonlinear density is subtracted on large scales as desired. In 2LPT 
this Lagrangian $F_2$ algorithm is equivalent to the Eulerian $F_2$
method (at the level of reconstructed densities, and up to the freedom which fields to smooth).

\section{Overview of Reconstruction Algorithms}
\label{se:RecOverview}

Before discussing simulations and numerical results, we briefly summarize all reconstruction algorithms in Table~\ref{tab:rec_overview} and in the following list.  The Eulerian algorithms that operate directly at the field level are:
\begin{enumerate}
\item \emph{EGS: Eulerian Growth-Shift Reconstruction (Section~\ref{se:EGS})}\\
Nonlinear growth and shift parts are subtracted from the unreconstructed density $\delta$,
\begin{equation}
  \label{eq:delta_EGS}
  \delta_\mathrm{EGS}^\rec(\vx) = \delta(\vx) - \delta_R(\vx)\delta(\vx)-\vs(\vx)\cdot\nabla\delta(\vx),
\end{equation}
where $\delta_R$ is the smoothed density and $\vs$ is defined in \eqq{vs_def2}. This follows from linearizing the velocity but not the mass density in the  nonlinear continuity equation.  The reconstructed power spectrum  combines the unreconstructed power and  specific $3$- and $4$-point statistics of the unreconstructed density, see \eqq{P_EGS}.
\item\emph{EF2: Eulerian F2 Reconstruction (Section~\ref{se:EF2Rec})}\\
Aiming to reverse the full second-order
  $\delta^{(2)}\sim F_2\delta*\delta$ part of the unreconstructed
  density, we form
\begin{equation}
  \label{eq:delta_F2rec}
  \delta_\mathrm{EF2}^\rec(\vx) =
  \delta(\vx)-\frac{17}{21}\delta_R(\vx)\delta(\vx)
  -\vs(\vx)\cdot\nabla\delta(\vx) -\frac{4}{21}K^2_R(\vx).
\end{equation}
This is similar to the growth-shift reconstruction \eq{delta_EGS} but an additional tidal $K^2_R$ part is subtracted and the growth term  has a slightly different coefficient.  The reconstructed power spectrum is again a combination of specific 2-, 3- and 4-point statistics of the unreconstructed density similar to \eqq{P_EGS}.
\item \emph{ERR: Eulerian random-random reconstruction}\\
This Eulerian algorithm resembles the Lagrangian random-random LRR algorithm (see \eqq{delta_LRR_2nd_order}),
\begin{equation}
  \label{eq:73}
\delta_\mathrm{ERR}^\rec=\delta(\vx) - \frac{2}{3}\delta_R^2(\vx)
- \vs(\vx)\cdot\nabla\delta_R(\vx) - \frac{1}{3}K_{RR}^2(\vx) .
\end{equation}
\end{enumerate}

\begin{table}[t]
\centering
\renewcommand{\arraystretch}{1.4}
\begin{ruledtabular}
\begin{tabular}{@{}lllll@{}}
& \phantom{abc} & \emph{Eulerian}  & \phantom{abc} & \emph{Lagrangian}
\\ 
\emph{Growth-Shift} && 
$\delta_\mathrm{EGS}^\rec(\vx)=\delta(\vx)-\delta_R(\vx)\delta(\vx)-\vs(\vx)\cdot\nabla\delta(\vx)$
&& $\delta_\mathrm{LGS}^\rec=\delta_d[\vs]-\delta_s[\vs]$
\\
\emph{Full F2} && 
$\delta_\mathrm{EF2}^\rec(\vx)=\delta(\vx)-\frac{17}{21}\delta_R(\vx)\delta(\vx) -\vs(\vx)\nabla\delta(\vx) -\frac{4}{21}K^2_R(\vx)$
&& $\delta_\mathrm{LF2}^\rec=\frac{3}{7}\delta_\mathrm{LGS}^\rec + \frac{4}{7} \delta_\mathrm{LRR}^\rec$\\
\emph{Random-Random} &&
$\delta_\mathrm{ERR}^\rec=\delta(\vx) - \frac{2}{3}\delta_R^2(\vx)- \vs(\vx)\cdot\nabla\delta_R(\vx) - \frac{1}{3}K_{RR}^2(\vx)$
&& $\delta_\mathrm{LRR}^\rec=\delta-\frac{1}{2}\left\{\delta_s[\vs]+\delta_s[-\vs]\right\}$ 
\end{tabular}
\end{ruledtabular}
\caption{Overview of the reconstruction algorithms considered in this paper. Lagrangian reconstructions involve
  displacements of objects which is not the case for Eulerian reconstructions. Methods in the same row agree at second order LPT.  The standard BAO reconstruction   algorithm is called Lagrangian growth-shift reconstruction and listed on the upper right. }
\label{tab:rec_overview}
\end{table}

Additionally, we consider the corresponding \emph{Lagrangian} reconstruction algorithms that require displacing objects in catalogs:
\begin{enumerate}
\item \emph{LGS: Lagrangian Growth-Shift Reconstruction (Section~\ref{se:LGSandLRR})}\\
  This is the standard BAO reconstruction algorithm proposed in \cite{EisensteinRecSims0604362}: Objects in clustered and random catalogs are displaced by the large-scale negative Zeldovich displacement field $\vs$, and the difference between the corresponding `displaced' and `shifted 'densities $\delta_d[\vs]$ and $\delta_s[\vs]$ gives the reconstructed density,
  \begin{eqnarray}
    \label{eq:delta_LGS_def}
    \delta_\mathrm{LGS}^\rec = \delta_d[\vs]-\delta_s[\vs].
  \end{eqnarray}
\item \emph{LF2: Lagrangian F2 Reconstruction (Section~\ref{se:LF2})}\\
The LGS and LRR methods are combined such that the second order $F_2$ part of the unreconstructed density is subtracted on large scales,
\begin{equation}
  \label{eq:delta_PF2rec_def1}
\delta_\mathrm{LF2}^\rec\equiv\frac{3}{7}\delta_\mathrm{LGS}^\rec
+ \frac{4}{7} \delta_\mathrm{LRR}^\rec.
\end{equation}
\item \emph{LRR: Lagrangian Random-Random reconstruction (Section~\ref{se:LGSandLRR})}\\
  This involves two oppositely shifted randoms: Randoms are shifted by the negative Zeldovich displacement, $\delta_s[\vs]$, and by the positive Zeldovich displacement, $\delta_s[-\vs]$, and their mean is subtracted from the observed density,
\begin{equation}
  \label{eq:delta_LRR}
  \delta_\mathrm{LRR}^\rec = \delta-\frac{1}{2}\left\{\delta_s[\vs]+\delta_s[-\vs]\right\}.
\end{equation}
The result agrees with the linear density at first order and suppresses the nonlinear growth and shift term at second order.
\end{enumerate}

As explained before, the 2LPT model for the reconstructed density of each Eulerian algorithm agrees with the corresponding Lagrangian-reconstructed density, although the algorithms are operationally very different.   All reconstruction algorithms leave the power spectrum unchanged on the very largest scales and reverse nonlinearities on smaller scales in different ways.

\section{Simulations}

To test the performance of the above reconstruction algorithms, we ran several N-body simulations with the \yucode code \cite{YuCode}, derived from the parallel COLA \cite{TassevCOLA} implementation of \cite{Koda1507,2014MNRAS.441.3524K}.  This is a particle-mesh (PM) code that uses linear time stepping in the scaling factor $a$ to produce accurate large scale structure at a fraction of the total computing time of a typical TreePM N-body simulation.  We set up 2LPT initial conditions for $2048^3$ DM particles in a box of length $L=1380\mathrm{Mpc}/h$ per side and evolve them from the initial redshift $z_i = 99$ to present time with $80$ time steps.  From $z=99$ to $z=1$ we use a $4096^3$ PM grid; at $z<1$ we switch to a $6144^3$ PM grid.  We only use a single snapshot at $z=0.55$, which is taken after 52 time steps. We run three independent realizations by generating Gaussian random fields following an initial DM power spectrum.  For each realization we also run a nowiggle simulation that has the same phases but is obtained from a nowiggle power spectrum where BAO wiggles are smoothed out.  The cosmic variance caused by broadband fluctuations can thus be cancelled when comparing wiggle and nowiggle simulations \cite{pacoWiggleNowiggleSims,Tobias1504.04366}.    We assume a flat $\Lambda$CDM cosmology with $\Omega_m=0.272, h=0.702, \sigma_8=0.807$ for the N-body simulations. The convergence of the simulations is discussed in Appendix~\ref{se:SimConvergence}.

For all plots that do not require wiggle and nowiggle simulations, we instead use `RunPB' N-body simulations produced by Martin White with the TreePM N-body code of \cite{mwhiteTreePM2002}.  These simulations were also used in \cite{beth1404,mwhite1408,marcel1411}.  They have 10 realizations of $2048^3$ DM particles in a box with side length $L=1380\mathrm{Mpc}/h$, and were started with 2LPT initial conditions at $z_i=75$ for an initial power spectrum with BAO wiggles (these simulations were not run for an initial nowiggle power spectrum).  The cosmology is flat $\Lambda\mathrm{CDM}$ with $\Omega_bh^2 = 0.022, \Omega_mh^2 = 0.139, n_s = 0.965, h = 0.69$ and $\sigma_8 = 0.82$, and we again only use one snapshot at $z=0.55$.

To speed up reconstructions we work with $1\%$ DM subsamples\footnote{For $R=15\mathrm{Mpc}/h$ smoothing, the relative performance of the reconstructions does not change qualitatively for $0.1\%$ and $0.01\%$ clustered subsamples, if the random catalogs for Lagrangian reconstructions contain as many particles as a $1\%$ clustered subsample.} of clustered catalogs, where the same random subset of particles is selected from wiggle and nowiggle simulations.  
Random catalogs are generated by distributing equally many particles randomly in the box.  
Densities are computed on $512^3$ grids using the cloud-in-cell (CIC) scheme. The CIC kernel
is deconvolved from the density grids.   For densities that need to be smoothed when entering in a reconstruction method we employ Gaussian smoothing with kernel $W_R(k)=\exp(-\tfrac{1}{2}k^2R^2)$.  Following \cite{Padmanabhan2012BAORec,AndersonBAODR9_1203} we choose $R=15\mathrm{Mpc}/h$ as the fiducial smoothing scale (see Appendix~\ref{se:VaryR} for different smoothing scales).

For Eulerian reconstructions, we calculate quadratic fields like $\vPsi(\vx)\cdot\nabla\delta(\vx)$ by Fourier-transforming $\delta(\vx)$ to $k$ space, multiply the result by appropriate filters like $\vk/k^2$, Fourier-transform the result back to $x$ space and multiply fields there.  These fields are then combined according to the particular reconstruction algorithm under consideration. Finally, we measure the power spectrum of the reconstructed density. For more details we refer the reader to \cite{marcel1411} where spectra of the same quadratic fields were measured.

For Lagrangian reconstructions, the displacement field $\vs$ is computed in $k$ space using \eqq{vs_def2} (see also \cite{Burden1504.02591}). This is then Fourier-transformed to $x$ space and linearly interpolated to the locations of particles in clustered and random catalogs. The particles are displaced according to the interpolated displacement field and  the density of the displaced catalogs is obtained with the CIC method. Finally we form the power spectrum of the reconstructed density.

To cancel the broadband cosmic variance between wiggle and nowiggle simulations when post-processing measured spectra for plots, we compute the required combination of spectra (e.g.~the fractional difference between wiggle and nowiggle simulations) for each realization individually, and only average the final result over realizations.   Error bars are standard errors of the mean obtained in this way, and estimated from the scatter between realizations.  Due to the cosmic variance cancellation these errors are typically very small and we will therefore omit them in most plots for clarity.

\section{Numerical Results}

We now discuss how well the reconstruction algorithms perform in simulations.  For clarity
we first discuss the performance of the Eulerian and Lagrangian
growth-shift reconstructions EGS and LGS in isolation and study in
detail where additional BAO information comes from.  This is then
followed by a more comprehensive comparison of all algorithms in
Section \ref{se:F2andRRNumerics}.  To be quantitative, we only consider power spectra here. Appendix~\ref{se:SliceComparisons} compares the algorithms at a somewhat more qualitative level using 2D density slices, showing that the density changes due to Eulerian and Lagrangian reconstructions are very similar and agree with 2LPT expectations at the field level.

\subsection{Growth-Shift Reconstructions}

\subsubsection{Overall Comparison}

\begin{figure}[tp]
\centerline{
\includegraphics[height=0.3\textheight]{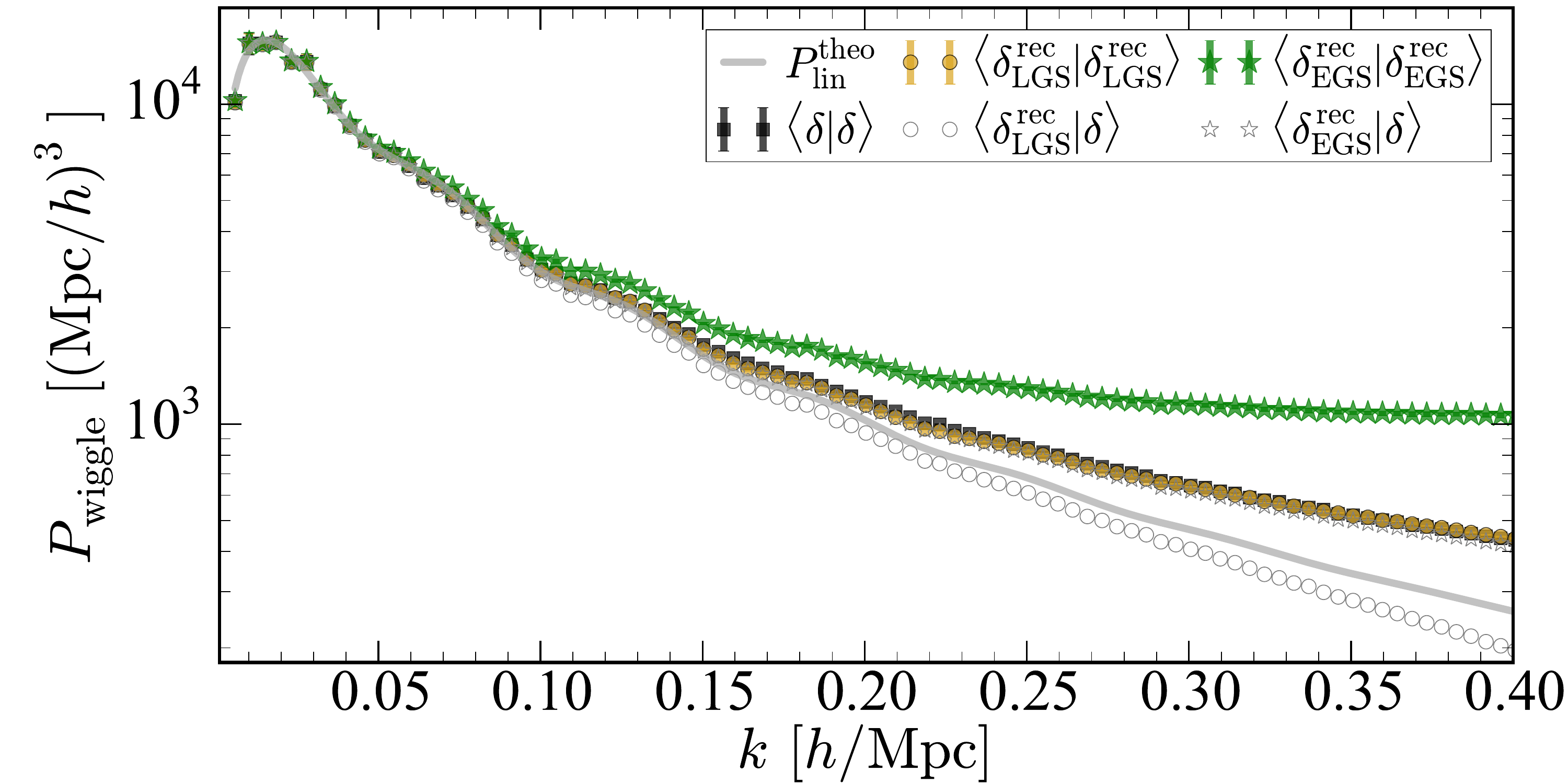}}
\caption{Total DM power spectra before reconstruction (black
  squares), after standard Lagrangian growth-shift LGS reconstruction
  by particle displacements (circles), and after Eulerian
  growth-shift EGS reconstruction based on combining 2-, 3- and 4-point
  information (stars), for $1\%$ subsamples of simulations with
  BAO wiggles.  Filled symbols show density auto-spectra, while open
  symbols show cross-spectra between the densities before and after
  reconstruction.  Results are averaged over 3 realizations and error
  bars correspond to the standard error of the mean (they are smaller
  than the symbols except at very low $k$ where their size
  is similar to that of the symbols).  }
\label{fig:GSBroadbandSpectra}
\end{figure}

\begin{figure}[tp]
\centerline{
\includegraphics[height=0.3\textheight]{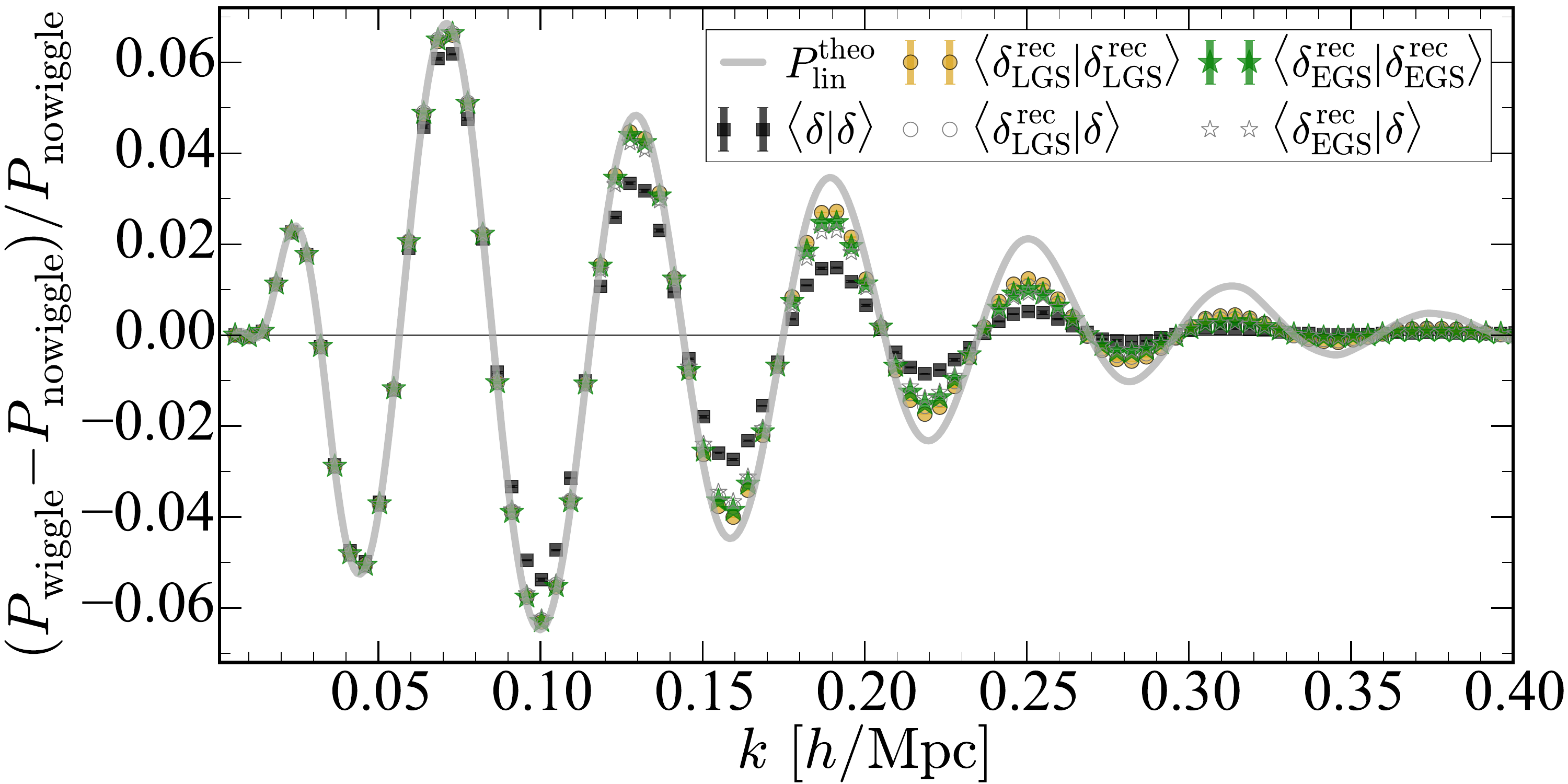}}
\caption{Illustration of the BAO signal: Fractional difference of the power spectra in Fig.~\ref{fig:GSBroadbandSpectra} between wiggle- and no-wiggle simulations. Again, error bars are smaller than the symbols.}
\label{fig:GSSignal}
\end{figure}

Fig.~\ref{fig:GSBroadbandSpectra} shows power spectra before and after
reconstruction.  For the Lagrangian LGS method, the broadband shape is
similar before and after reconstruction, whereas the Eulerian EGS reconstruction adds
additional broadband power at high $k$.  This is only true for the particular smoothing that we use for the EGS method. E.g.~if we used symmetric smoothing where all fields entering into quadratic fields are smoothed rather than just one of them, then the broadband of the reconstructed power agrees with that before reconstruction. However, symmetric smoothing yields significantly lower BAO signal-to-noise than asymmetric smoothing, which makes sense because asymmetric smoothing allows small-scale fluctuations to be shifted by large-scale flows, whereas symmetric smoothing erases all small-scale information. 

To highlight the BAO signal per $k$-mode, Fig.~\ref{fig:GSSignal} shows the
fractional difference of spectra between simulations with and without
wiggles. After reconstruction, this fractional
difference has more pronounced wiggles than before reconstruction and
is closer to linear theory, implying an enhanced BAO signal.   The
enhancement is  similar for Eulerian and Lagrangian
reconstructions.   Most (but not all) of the signal can also be obtained by forming
the cross-spectrum between reconstructed and unreconstructed
densities, as shown by non-filled symbols in Fig.~\ref{fig:GSSignal}.

\begin{figure}[tp]
\centerline{
\includegraphics[height=0.3\textheight]{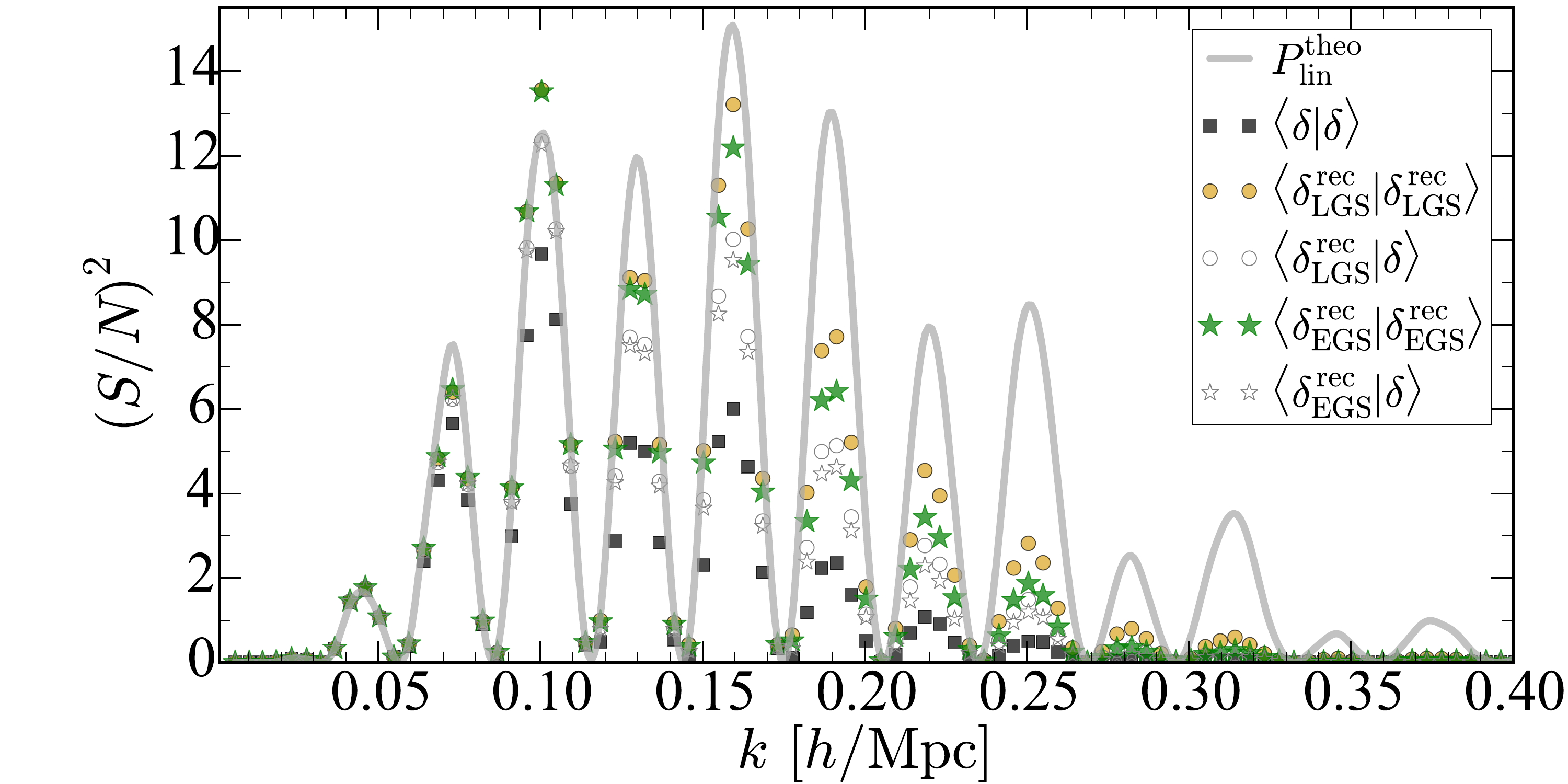}}
\caption{Signal-to-noise-squared for the BAO wiggles as a function of wavenumber $k$. The signal is given by the difference of the spectra specified in the legend between wiggle and nowiggle simulations, $S=P_\mathrm{wiggle}-P_\mathrm{nowiggle}$. The squared noise is approximated by \eqq{GaussianNoise}.  Error bars are not shown for clarity.} 
\label{fig:GSSNsq}
\end{figure}

Fig.~\ref{fig:GSSNsq} shows the signal-to-noise-squared per $k$ mode estimated by
assuming that the BAO signal is given by the difference of
power spectra between wiggle and nowiggle simulations, and
approximating the covariance with the Gaussian expectation
\begin{equation}
  \label{eq:GaussianNoise}
\mathrm{cov} (\hat P(k), \hat P(k')) = \delta_{k,k'}\frac{2}{N_\mathrm{modes}(k)}\langle\hat P(k)\rangle^2,
\end{equation}
where the number of modes per $k$-bin and the power spectrum
expectation value are computed from the simulations; see Appendix~\ref{se:GaussianCov} for justification of this approximation.
Fig.~\ref{fig:GSSNsq} shows
that the BAO signal-to-noise-squared is clearly enhanced after
EGS and LGS reconstructions. The algorithms perform equally well for $k\le 0.15h/\mathrm{Mpc}$.  On smaller scales, the Lagrangian LGS algorithm performs slightly
better than the Eulerian EGS algorithm.  This is caused by the larger broadband noise of the EGS algorithm in this regime (see Fig.~\ref{fig:GSBroadbandSpectra}).

\begin{figure}[tp]
\centerline{
\includegraphics[height=0.3\textheight]{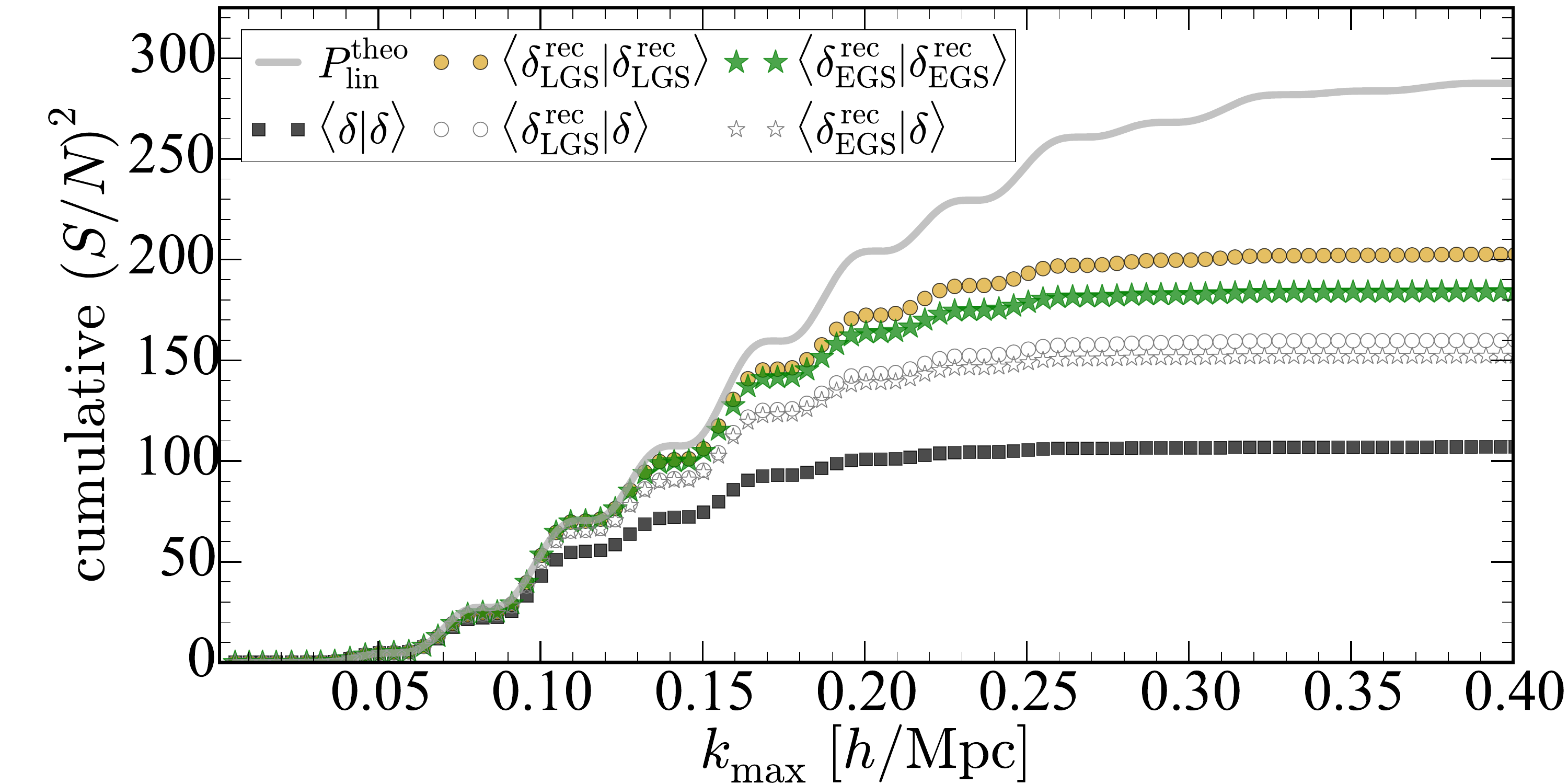}}
\caption{Cumulative BAO signal-to-noise-squared as a function of $k_\mathrm{max}$ for EGS and LGS reconstruction.}
\label{fig:GScumSNsq}
\end{figure}

To better quantify the overall performance, Fig.~\ref{fig:GScumSNsq}
shows the cumulative BAO signal-to-noise-squared as a function of $k_\mathrm{max}$.  The
Lagrangian LGS reconstruction improves the BAO signal-to-noise by a
factor $1.4$ for $k_\mathrm{max}\sim 0.4h/\mathrm{Mpc}$. The
Eulerian EGS algorithm yields $\sqrt{184/202}\sim 95\%$ of the BAO
signal-to-noise of the LGS method. Thus, Eulerian and Lagrangian growth-shift
reconstructions perform very similarly, differing by less than $5\%$ in their
total BAO signal-to-noise (see also Table~\ref{tab:SNsummary} later).  

Using cross-spectra between reconstructed and unreconstructed
densities instead of auto-spectra of reconstructed densities degrades
the cumulative signal-to-noise by $\sim 5\%$ for each of the two
reconstruction methods.

\subsubsection{3-point vs 4-point}
\label{se:3ptOr4pt}
Does most of the additional BAO information gained by reconstruction come from the 3-point or the 4-point function of the unreconstructed density? To answer this, we write the reconstructed density as
\begin{equation}
  \label{eq:72}
 \delta^\mathrm{rec}=\delta+(\delta^\mathrm{rec}-\delta),
\end{equation}
where $\delta$ is the unreconstructed density.   For Eulerian reconstructions, the density change due to reconstruction is quadratic in the unreconstructed density, $(\delta_\mathrm{rec}-\delta)\sim\mathcal{O}(\delta^2)$.  The reconstructed power spectrum therefore splits into 2-, 3- and 4-point parts as
\begin{equation}
  \label{eq:EulerRecRecSplit}
  \langle\delta^\mathrm{rec}|\delta^\mathrm{rec}\rangle
= \underbrace{\la\delta|\delta\ra}_\mathrm{2pt} 
+ \underbrace{2 \la(\delta^\mathrm{rec}-\delta)|\delta\ra}_\mathrm{3pt}
+\underbrace{\la(\delta^\mathrm{rec}-\delta)|(\delta^\mathrm{rec}-\delta)\ra}_\mathrm{4pt}.
\end{equation}
For example, the cross-spectrum between $(\delta_\mathrm{rec}-\delta)$ and $\delta$ is a cross-spectrum between a field quadratic in $\delta$ and $\delta$ itself, and is therefore given by a particularly integrated bispectrum of $\delta$ \cite{marcel1411}. 

\begin{figure}[tp]
\centerline{
\includegraphics[height=0.3\textheight]{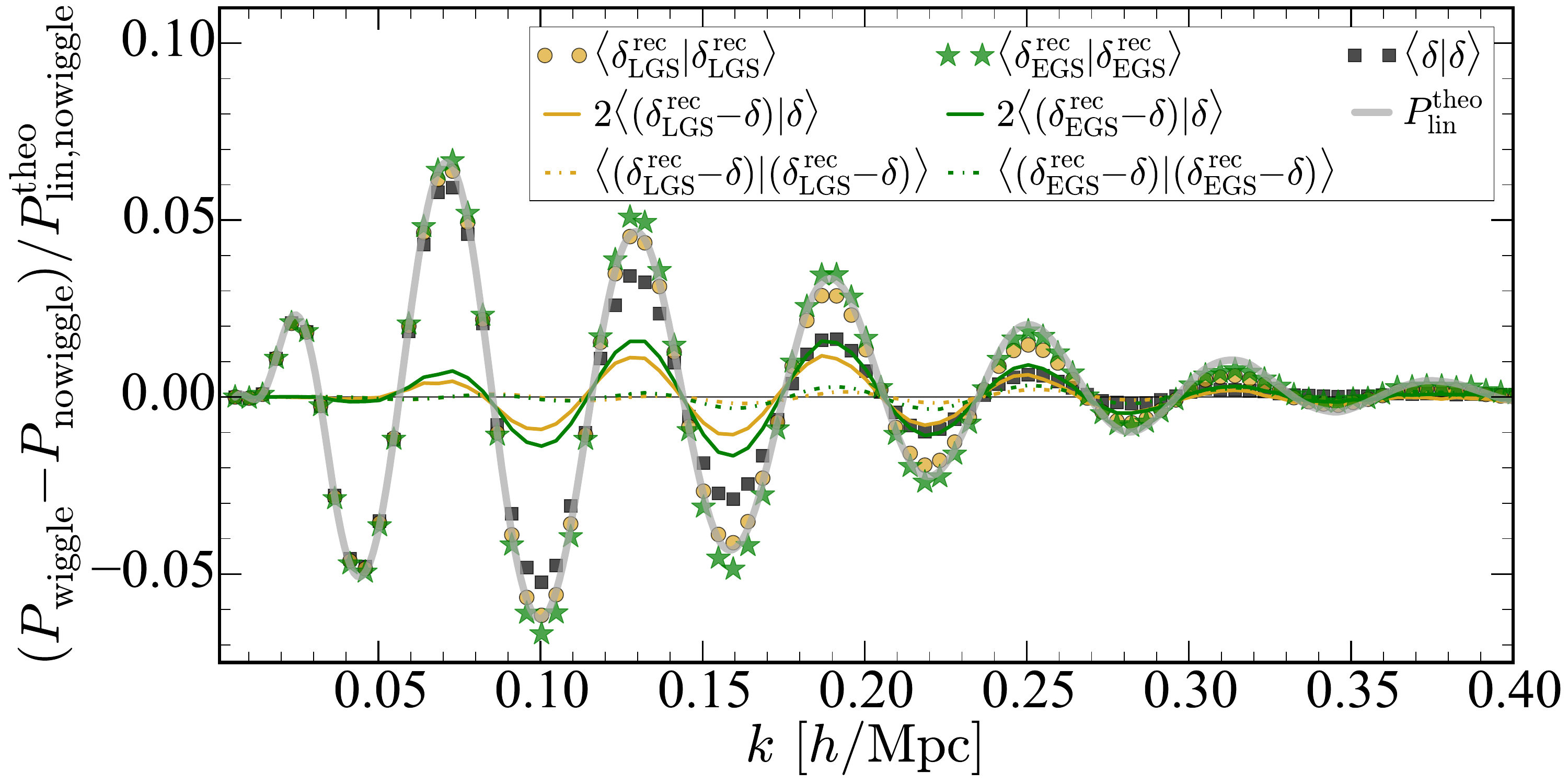}}
\caption{Illustration of 3- and 4-point contributions to reconstructed power spectra.  The full reconstructed power spectra (circles and stars) can be obtained by adding the 3-point part (thin solid) and 4-point part (dash-dotted) to the pre-reconstruction power spectrum (black squares);  see Section~\ref{se:3ptOr4pt} for details.  The curves show differences of the spectra specified in the legend between wiggle and nowiggle simulations, divided by the theoretical linear no-wiggle density power spectrum.  At all $k$ error bars estimated from the scatter between the 3 realizations are roughly the same size as the thickness of the colored curves and smaller than the size of the circle, star and square symbols; they are not shown for clarity.}
\label{fig:Rec3ptvs4pt}
\end{figure}

Simulation measurements of the terms on the right hand side of \eqq{EulerRecRecSplit} are shown
in Fig.~\ref{fig:Rec3ptvs4pt}.  For EGS reconstruction, the enhanced
BAO signal is almost entirely due to the 3-point part of the
signal (thin solid), whereas the additional signal from the 4-point
(dash-dotted) is almost negligibly small.

For Lagrangian reconstructions the difference between reconstructed
and unreconstructed densities is in general not simply quadratic in
the observed density any more, because the reconstructed density is
obtained by calculating new densities from shifted and displaced
catalogs that cannot easily be related to the density of the original
catalog.  It is therefore less transparent to study the role of 3- and
4-point signal in this case.  Still, modeling Lagrangian
reconstruction from a Lagrangian theory perspective shows that
reconstructed and unreconstructed densities agree at first order in
the linear density, so that the leading-order difference between
reconstructed and unreconstructed densities is quadratic in the linear
density; see \eqq{delta_LGS_x_LPTfinal} and
Appendix~\ref{se:LagrangianTheory}.  The spectra on the right hand
side of \eqq{EulerRecRecSplit} are therefore still related to 3- and
4-point contributions if third order corrections are ignored.
Fig.~\ref{fig:Rec3ptvs4pt} shows that for LGS reconstruction these
spectra are similar to EGS reconstruction, indicating that the
enhanced BAO signal from LGS reconstruction is also mostly due to the effectively
included 3-point signal.

\subsubsection{13 and 22 contributions to the 3-point part of the
  reconstructed power spectrum}
\label{se:13_22_split}

\begin{figure}[tp]
\centerline{
\includegraphics[height=0.3\textheight]{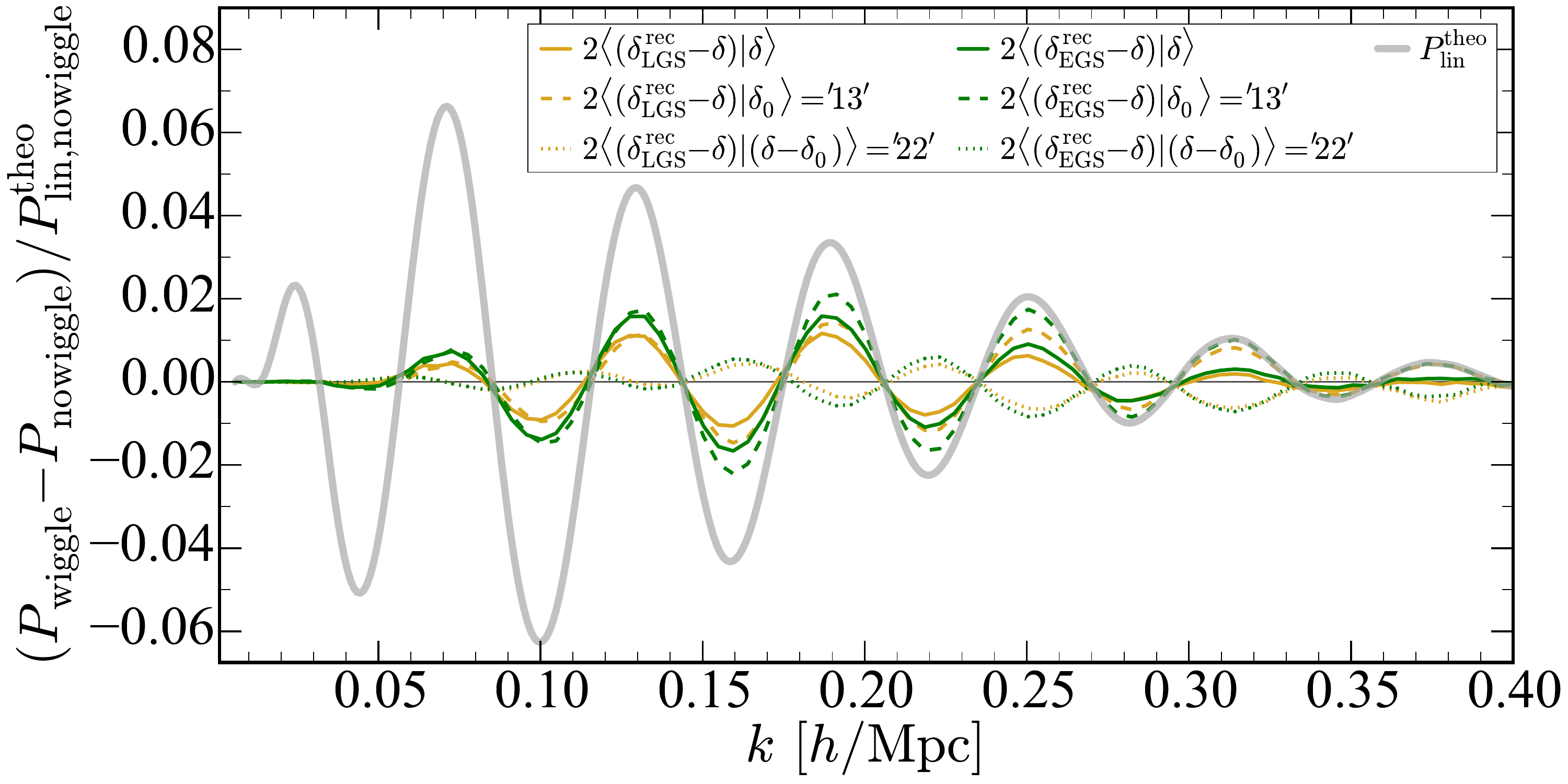}}
\caption{Separation of the reconstruction 3-point part $2\la(\delta^\rec-\delta)|\delta\ra$ (thin solid) from Fig.~\ref{fig:Rec3ptvs4pt} into `13' contribution (dashed) and `22' contribution (dotted) according to \eqq{13_22_split}, measured from simulations for standard LGS reconstruction (yellow) and EGS reconstruction (green).  The linear density $\delta_0$ in the cross-spectra is approximated by the density of the simulation initial conditions, rescaled to $z=0.55$ with the linear growth factor.  Adding dotted to dashed curves gives the thin solid curve of the same color.   The curves show differences of the spectra specified in the legend between wiggle and nowiggle simulations, divided by the theoretical linear no-wiggle density power spectrum.  Error bars of all colored curves estimated from the scatter between the 3 realizations are roughly the same size as the thickness of the curves at all $k$; they are not shown for clarity.}
\label{fig:spectra_13_22_splits_wiggle_m_nowiggle_over_Plintheo}
\end{figure}

\begin{figure}[tp]
\centerline{
\includegraphics[width=0.5\textwidth]{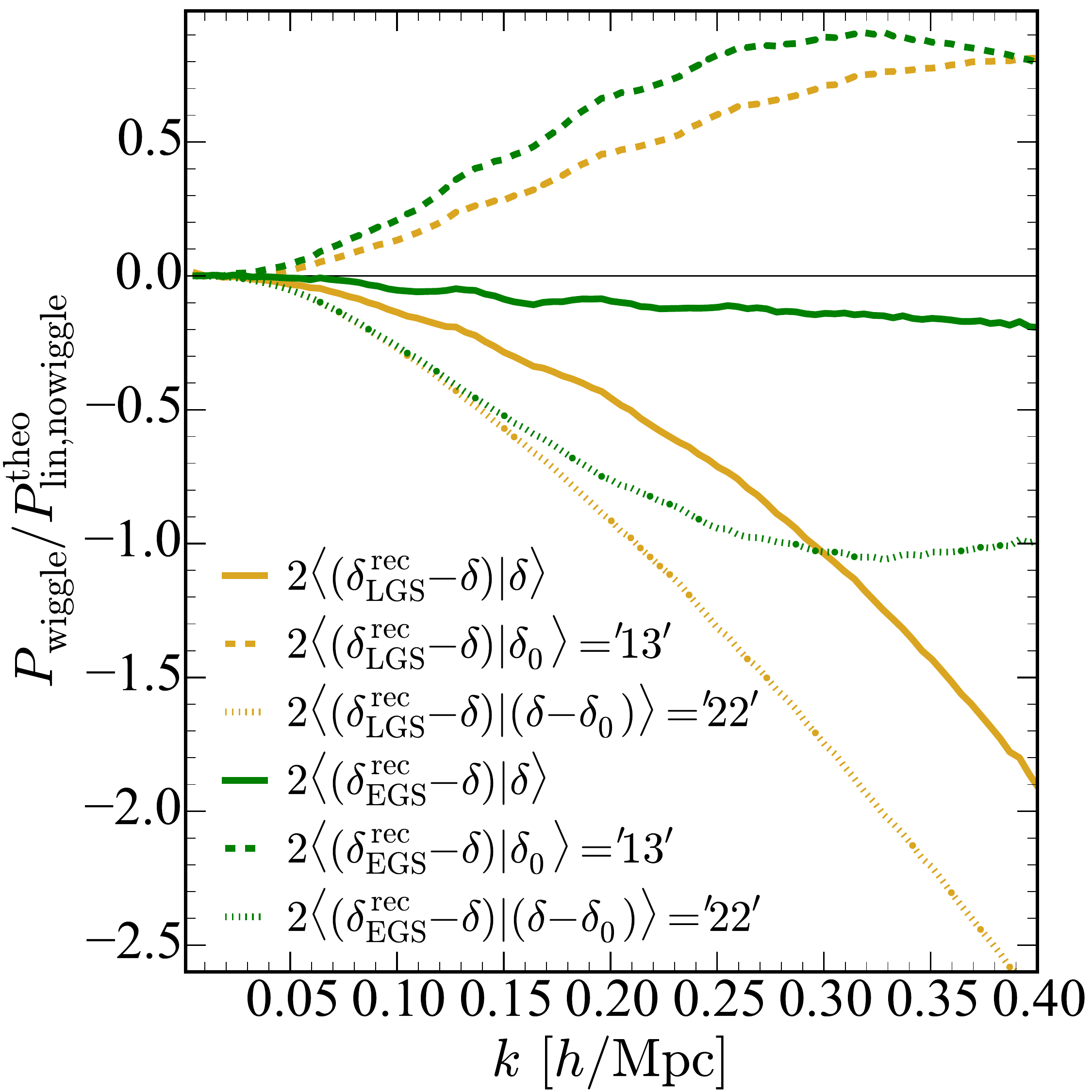}}
\caption{Same plot as before but including broadband information by plotting spectra specified in the legend in wiggle simulations divided by the theoretical linear no-wiggle density power spectrum. Error bars are roughly the same size as the thickness of the curves at $k\lesssim 0.15h/\mathrm{Mpc}$ and twice that size at higher $k$, but they are not shown for clarity. The plot shows the average over the 10 RunPB N-body realizations available to us. For the 3 \yucode realizations that we ran for the previous plots this looks very similar but slightly noisier because of the smaller number of realizations.   }
\label{fig:spectra_13_22_splits}
\end{figure}

To further elucidate the origin of the enhanced BAO signal after
reconstruction we split the observed density $\delta$ into linear
and nonlinear part
\begin{equation}
  \label{eq:63}
  \delta = \delta_0 + (\delta-\delta_0),
\end{equation}
where $\delta_0$ is the linear density.  Then, the 3-point-like term
from the
last section splits into
\begin{equation}
  \label{eq:13_22_split}
  \langle(\delta^\mathrm{rec}-\delta)|\delta\rangle
=
\underbrace{\langle(\delta^\mathrm{rec}-\delta)|\delta_0\rangle}_{\sim\la\Delta^{(3)}\delta_0\ra+\mathcal{O}(\delta_0^6)}
+
\underbrace{\langle(\delta^\mathrm{rec}-\delta)|(\delta-\delta_0)\rangle}_{\sim\la\Delta^{(2)}\delta^{(2)}\ra+\mathcal{O}(\delta_0^6)}.
\end{equation}
Expanding
\begin{equation}
  \label{eq:46}
 \delta^\mathrm{rec}-\delta=\Delta^{(2)}+\Delta^{(3)}+\cdots
\end{equation}
where $\Delta^{(n)}$ is n-th order in the linear Gaussian
density $\delta_0$, implies that the first term on the right hand side
of \eqq{13_22_split} corresponds to a contraction of $\Delta^{(3)}$
with $\delta_0$ at leading order in $\delta_0$; we call this the `13'
contribution.  The second term on the right hand side of
\eqq{13_22_split} corresponds to a `22' contribution at leading order.

To see this more explicitly, we expand the field change due to
reconstruction up to third order in $\delta_0$,
\begin{align}
  \label{eq:70}
  \delta^\mathrm{rec}-\delta
  = &\; \Delta^{(2)}+\Delta^{(3)}\\
 = &\; \int_{\vk_1,\vk_2}
  (2\pi)^3\delta_D(\vk-\vk_1-\vk_2) D_2^{(s)}(\vk_1,\vk_2)
  \delta_0(\vk_1)\delta_0(\vk_2)
\\
&\;+\int_{\vk_1,\vk_2,\vk_3}
  (2\pi)^3\delta_D(\vk-\vk_1-\vk_2-\vk_3) D_3^{(s)}(\vk_1,\vk_2,\vk_3)
  \delta_0(\vk_1)\delta_0(\vk_2)\delta_0(\vk_3),
\end{align}
where $D_2^{(s)}$ and $D_3^{(s)}$ are symmetric kernels whose explicit form
depends on the reconstruction algorithm under consideration. Then,
the 13 contribution to \eqq{13_22_split} becomes
\begin{align}
  \label{eq:Deltadelta13}
  \la\Delta^{(3)}(\vk)\delta_0(\vk')\ra
=
3\,(2\pi)^3\delta_D(\vk+\vk')\,P_\mathrm{lin}(k)\,
\int_{\vk_1}D_3^{(s)}(\vk_1,-\vk_1,\vk)P_\mathrm{lin}(k_1),
\end{align}
and similarly the 22 contribution is
\begin{align}
  \label{eq:Deltadelta22}
  \la\Delta^{(2)}(\vk)\delta^{(2)}(\vk')\ra
=
2\,(2\pi)^3\delta_D(\vk+\vk')\,
\int_{\vk_1,\vk_2} (2\pi)^3\delta_D(\vk-\vk_1-\vk_2)\,
 D_2^{(s)}(\vk_1,\vk_2)
 F_2(\vk_1,\vk_2)
\,P_\mathrm{lin}(k_1)P_\mathrm{lin}(k_2).
\end{align}

Figures~\ref{fig:spectra_13_22_splits_wiggle_m_nowiggle_over_Plintheo} and \ref{fig:spectra_13_22_splits} show the spectra on the right
hand side of \eqq{13_22_split} measured from simulations.  The linear density $\delta_0$ is approximated by estimating the density
of the initial conditions of the simulations and rescaling it to $z=0.55$ with the linear growth factor.  The measured 13 term
has pronounced BAO wiggles, which can be attributed to the fact that
the theoretical 13 term of \eqq{Deltadelta13} is proportional to
$P_\mathrm{lin}(k)$ which has BAO wiggles.\footnote{This makes the
  approximation that the convolution of $P_\mathrm{lin}$ with $D_3$ in
  \eqq{Deltadelta13} yields a broadband power that affects the
  wiggles only in a sub-dominant way.  Also, note that
  even at sixth or higher order in $\delta_0$, the $\delta_0(\vk')$
  factor on the l.h.s.~implies that \eqq{Deltadelta13} will always be
  of the form $P_\mathrm{lin}(k)$ times some integral over weighted
  power spectra, which can lead to further enhancements of the BAO
  wiggles.}  
In contrast, the measured 22 contribution in Fig.~\ref{fig:spectra_13_22_splits_wiggle_m_nowiggle_over_Plintheo} oscillates out of phase with the linear BAO wiggles, which shifts the BAO scale inferred from scales $k\lesssim 0.15h/\mathrm{Mpc}$.  This can be understood perturbatively from the form of the theoretical 22 expression in \eqq{Deltadelta22} \cite{padmanabhan0812,NikhilMartin0906,NohPadmanabhanWhite2009,SherwinZaldarriaga2012}.  At higher $k$, the 22 contribution is opposite in phase to the linear BAO wiggles and thus reduces the enhancement of BAO wiggles from the 13 contribution.\footnote{If the kernels were known, BAO information from the convoluted   22 contributions could still be extracted, but in practice the kernels are only known in
  perturbation theory which starts to break down on scales where
  reconstruction helps most, so that more information can be extracted
  by the standard approach of fitting an oscillatory function on top of a
  broadband shape out to smaller scales.}  

In summary, most of the additional BAO signal from reconstruction is due to the 13-type part of the 3-point contribution to the reconstructed power spectrum, whereas the 22-type 3-point and the 4-point do not enhance the signal much.

Figures~\ref{fig:spectra_13_22_splits_wiggle_m_nowiggle_over_Plintheo} and \ref{fig:spectra_13_22_splits} also compare the 13 and 22
contributions between the Eulerian and Lagrangian growth-shift
reconstruction algorithms. These algorithms have the same 22
contributions at $k\lesssim 0.15h/\mathrm{Mpc}$, which is expected
because they are equivalent when modeled with 2LPT (see
Appendix~\ref{se:LagrangianTheory}).  In contrast, the 13 contributions
differ even at low $k$, which indicates that the algorithms differ at
third order in the linear density.  This is not surprising since they
involve rather different transformations of the unreconstructed density.  Since the 13 contribution is most important for improving the BAO signal it may be worth considering third order reconstruction algorithms that subtract $F_3\delta*\delta*\delta$ correctly. However, this goes beyond the scope of this paper, and it should be noted that for Eulerian reconstructions such an extension might suffer from shot noise issues because e.g.~$\delta^3(\vx)$ is very sensitive to the largest peaks of the density.  In this case, the standard reconstruction method may be a more elegant solution, because it successfully reverses nonlinear BAO smoothing beyond second order without the need to cube any Eulerian fields.

All spectra in Fig.~\ref{fig:spectra_13_22_splits}  vanish on
very large scales ($k\lesssim 0.05h/\mathrm{Mpc}$) as expected because
both reconstructed and unreconstructed densities approach the linear
density in this regime.

\subsubsection{Growth vs Shift Term}

\begin{figure}[tp]
\centerline{
\includegraphics[height=0.3\textheight]{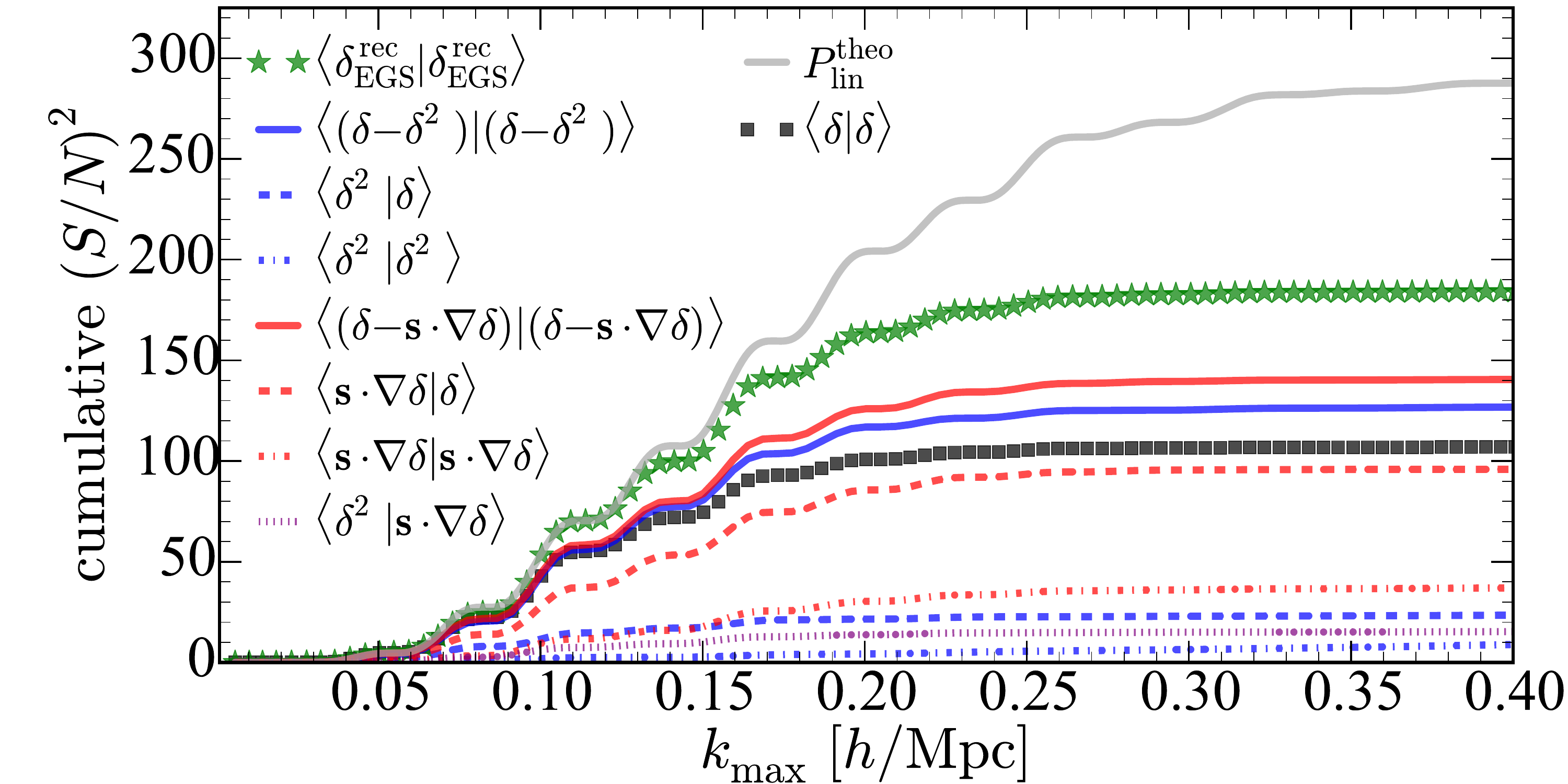}}
\caption{Separation of EGS reconstruction in contributions from growth term $\delta^2$ (blue) and shift term $\vs\cdot\nabla\delta$ (red), see \eqq{P_EGS}. The plot shows the cumulative BAO signal-to-noise-squared of spectra specified by the legend. Green stars show full EGS reconstruction, colored solid show reconstructions if only the growth (blue) or only the shift term (red) is used. Dashed lines show the signal-to-noise-squared of individual 3-point contributions to the full EGS reconstruction.   Dash-dotted and dotted show 4-point contributions.}
\label{fig:spectra_cumSNsq_split_in_delta2_and_shift}
\end{figure}

For Eulerian growth-shift reconstruction, $\delta_\mathrm{rec}^\mathrm{EGS}=\delta-\delta^2-\vs\cdot\nabla\delta$, the power spectrum of the reconstructed density can be split into six cross-spectra between unreconstructed density $\delta$, growth term $\delta^2$ and shift term $\vs\cdot\nabla\delta$, see \eqq{P_EGS}.  Fig.~\ref{fig:spectra_cumSNsq_split_in_delta2_and_shift} shows the cumulative BAO signal-to-noise from each of these contributions.\footnote{The individual signal-to-noise-squared curves in Fig.~\ref{fig:spectra_cumSNsq_split_in_delta2_and_shift} cannot easily be combined by eye because there are nontrivial correlations between the spectra. These are taken into account for the full reconstructed signal-to-noise-squared because spectra are combined realization by realization (which is equivalent to combining $\delta$ and the growth and shift term at the field level for each realization and then taking spectra of the resulting field).}    The cross-spectrum between shift term and the density is most important, which makes sense intuitively because this term corresponds to shifting back large-scale flows.   If the reconstruction only included the shift term, $\delta_\mathrm{rec}=\delta-\vs\cdot\nabla\delta$ (red solid in Fig.~\ref{fig:spectra_cumSNsq_split_in_delta2_and_shift}), then the improvement over performing no reconstruction (black squares) is however only half of the improvement obtained from the full EGS reconstruction (green stars).  Thus, the growth term is subdominant but adds very significant signal-to-noise so that it should not be omitted.  Also, if the growth term was not included, the reconstructed power spectrum would not yield the linear power spectrum on large scales, so that such a shift-only reconstruction scheme should not be taken seriously in any case.

\subsection{All Reconstructions}
\label{se:F2andRRNumerics}

So far we only discussed numerical results for the growth-shift reconstructions.  We will now turn to the F2 and random-random reconstructions introduced in Sections~\ref{se:EF2Rec} and \ref{se:LagrangianRecs}, and compare all reconstruction algorithms.

\begin{figure}[tp]
\centerline{
\includegraphics[height=0.3\textheight]{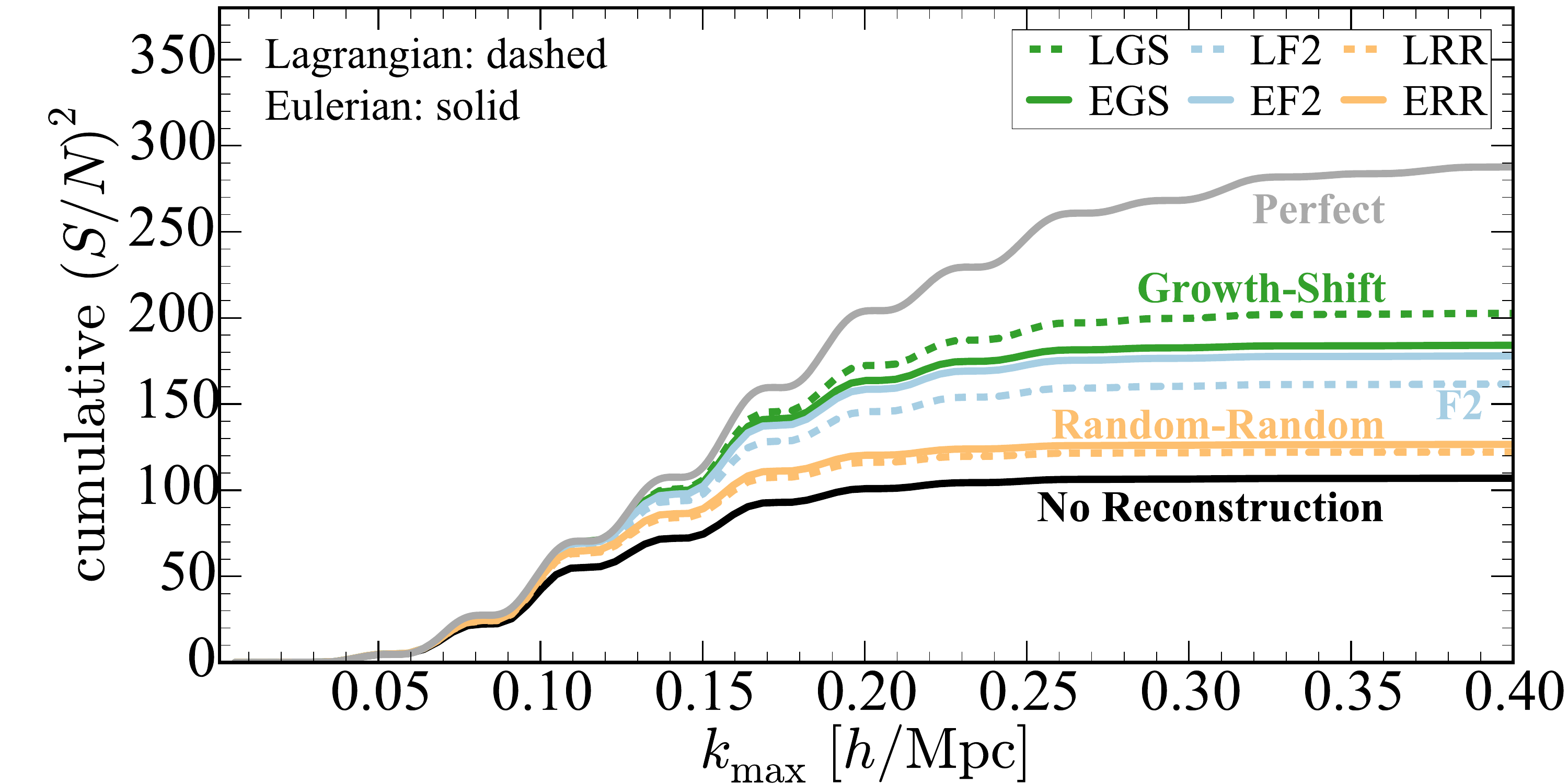}}
\caption{Cumulative BAO signal-to-noise-squared for reconstructed power spectra as a function of $k_\mathrm{max}$ measured from simulations for three Eulerian reconstruction algorithms (dashed colored) and the corresponding three Eulerian reconstruction algorithms (solid colored).  See Section~\ref{se:RecOverview} for a summary of the algorithms.  The signal-to-noise-squared of the density before reconstruction (black) and of the linear density (gray) are included for comparison.}
\label{fig:SixcumSNsq}
\end{figure}

All six reconstruction methods are compared in Fig.~\ref{fig:SixcumSNsq} in terms of their cumulative BAO signal-to-noise-squared as a function of $k_\mathrm{max}$.  Table~\ref{tab:SNsummary} shows the total BAO signal-to-noise up to $k_\mathrm{max}=0.4h/\mathrm{Mpc}$.   Overall, the standard Lagrangian growth-shift (LGS) algorithm works best.  Compared against that, the Eulerian growth-shift (EGS) and Eulerian F2 (EF2) methods perform almost as well, missing only $5\%$ and $6\%$ of the total BAO signal-to-noise, respectively.  The Lagrangian F2 (LF2) method turns out slightly worse, missing $11\%$ of the LGS signal-to-noise.   Both random-random reconstructions perform significantly worse, missing more than $20\%$ of the LGS signal-to-noise.  Note that the growth-shift and F2 algorithms perform equally well on large and intermediate scales, $k\le 0.15h/\mathrm{Mpc}$, as expected from 2LPT modeling, and they only begin to differ on smaller scales.  Thus, EGS, EF2 and LF2 only perform slightly worse than LGS because they partially miss highly nonlinear BAO information that the LGS algorithm restores slightly better.

The EGS and EF2 methods may be useful for applications because they perform almost as well as the standard LGS algorithm and have potential advantages due to their Eulerian nature (e.g.~they can be expressed as combinations of 2-, 3- and 4-point functions of the unreconstructed density).   The LF2 method and the random-random methods are not competitive in terms of signal-to-noise so they should not be used.  The performance of ERR can be improved by using asymmetric smoothing where only one of the fields entering quadratic fields is smoothed; in this case ERR performs almost as well as EF2 but still slightly worse.  For the LRR method it is less obvious how to reduce the smoothing to improve it, and we do not investigate this further.

\changed{So far we focused on the BAO signal-to-noise to assess reconstruction performance.  Another important aspect of the standard LGS reconstruction is that it removes a shift of the BAO scale induced by nonlinear clustering \cite{SeoBAOShiftPaper1,SeoBAOShift}.   We investigate this effect for our new algorithms as follows. For every estimated power spectrum we construct the fractional difference $(P_\mathrm{wiggle}-P_\mathrm{nowiggle})/P_\mathrm{nowiggle}$  shown in Fig.~\ref{fig:GSSignal}. Then we fit this with the damped shifted linear prediction $O(k/\alpha)\exp[-k^2\Sigma^2/2]$ for varying $\alpha$ and $\Sigma$, where  $O(k)\equiv [P^\mathrm{lin}_\mathrm{wiggle}(k)-P^\mathrm{lin}_\mathrm{nowiggle}(k)]/P^\mathrm{lin}_\mathrm{nowiggle}(k)$ corresponds to the gray line in Fig.~\ref{fig:GSSignal}.  Focusing on the BAO wiggles, we fit over the range $0.083\,h/\mathrm{Mpc} \leq k \leq 0.3\,h/\mathrm{Mpc}$.  Since the sample standard deviations estimated from the scatter between three realizations are somewhat uncertain for individual $k$ bins,
and since errors have no broad-band scale dependence because broad-band cosmic variance is cancelled,  we replace sample errors by their average over all $k$ bins, which gives a constant error of $2\times 10^{-4}$.  

Marginalizing over the damping parameter $\Sigma$, the best-fit values for the BAO shift $\alpha-1$ after LGS, EGS, LF2, EF2, LRR and ERR reconstructions are 0.003$\%$, 0.01$\%$, 0.06$\%$, 0.03$\%$, 0.13$\%$ and  0.13$\%$, respectively, and 0.25$\%$ without reconstruction. The 1$\sigma$ fitting uncertainty is roughly 0.01$\%$.  This shows that mitigation of the BAO shift is more efficient for reconstructions that also yield more BAO signal-to-noise. In particular, for LGS, EGS and LF2 reconstructions the residual BAO shift is at most $0.03\%$, which is negligibly small for upcoming experiments. Within the fitting uncertainty, the LGS and EGS reconstruction do not show evidence for any residual BAO shift. }

Our results are based on simulations and it is not clear how to explain them within the framework of second order perturbation theory where all methods agree up to smoothing and should thus perform similarly well.  The difference in performance could be attributed to different smoothing operations, e.g.~theory expressions for the random-random methods involve only the smoothed density, whereas theory expressions for the other reconstruction methods involve also unsmoothed fields so that e.g.~small-scale modes are shifted by large-scale smoothed displacements, which can yield improved reconstructions.   However this does not explain the (small) differences between the LGS, EGS, LF2 and EF2 algorithms.  These could be due to higher order corrections in perturbation theory, which we have not included in our modeling where the density was truncated at second order in the linear density.  The fact that the 13 part of the 3-point contribution to the reconstructed power spectrum seems to be most important in simulations (see Fig.~\ref{fig:spectra_13_22_splits_wiggle_m_nowiggle_over_Plintheo}) suggests that third-order corrections could be responsible for the slightly different performances of the LGS, EGS, LF2 and EF2 algorithms.  We do not investigate the performance differences of these algorithm in more detail here because the LGS, EGS and EF2 algorithms perform almost equally well.

\begin{table}[tp]
\centering
\renewcommand{\arraystretch}{1.4}
\begin{ruledtabular}
\begin{tabular}{@{}p{4.25cm}lrrrrrrrrrrrr@{}}
& \phantom{} & \multicolumn{2}{l}{Growth-Shift} &\phantom{} & \multicolumn{2}{l}{F2 reconstruction} & \phantom{} & \multicolumn{2}{l}{Random-Random} & \phantom{} &  & \phantom{} &  \\
\textbf{Reconstruction method} && LGS & EGS && LF2 & EF2 && LRR & ERR && Perfect && NoRec  \\
\textbf{BAO signal-to-noise} &&  14.2 &  13.6 && 12.7 &  13.3 &&  11.1 &  11.3 &&  17.0 &&  10.3\\
\textbf{Compared against LGS} &&  $\pm 0\,\%$ &  $-4.7\,\%$ && $-11\,\%$ &  $-6.3\,\%$ &&  $-22\,\%$ & $-21\,\%$ &&  $+19\,\%$ && $-27\,\%$ \\
\textbf{Compared against NoRec} && $+38\,\%$ &  $+31\,\%$ &&  $+23\,\%$ &  $+29\,\%$ && $+6.9\,\%$ &   $+8.8 \,\%$ &&  $+64\,\%$ &&   $\pm 0\,\%$ \\
\end{tabular}
\end{ruledtabular}
\caption{Total BAO signal-to-noise for $k_\mathrm{max}=0.4h/\mathrm{Mpc}$ for various reconstruction algorithms (obtained from Fig.~\ref{fig:SixcumSNsq}, based on simulations). `Perfect' refers to the BAO signal-to-noise of the linear density and `NoRec' to the BAO signal-to-noise of the measured nonlinear density without performing any reconstruction.  The second-to-last row shows how much of the signal-to-noise is lost compared to performing the standard LGS reconstruction.  The bottom row shows how much signal-to-noise is gained by reconstructions compared to performing no reconstruction. }
\label{tab:SNsummary}
\end{table}

\section{Conclusions}

Density reconstruction plays an important role in modern cosmology by significantly improving BAO information from galaxy surveys.  In this paper we present three new Eulerian reconstruction algorithms that operate directly on the pre-reconstruction density field, and two new Lagrangian reconstructions that displace objects in catalogs.  These algorithms are physically and operationally different from the conventionally used Lagrangian growth-shift (LGS) reconstruction.

The Eulerian growth-shift (EGS) reconstruction  algorithm is derived by reversing the nonlinear continuity equation, keeping the mass density fully nonperturbative and only approximating the velocity density in terms of its linear relation to the mass density.  The resulting reconstructed power spectrum is given by a particular combination of simple 2-, 3- and 4-point statistics of the unreconstructed density.  In our simulation setup, this combination yields the same BAO signal-to-noise as the standard LGS algorithm up to $k_\mathrm{max}=0.15h/\mathrm{Mpc}$, and $95\%$ of the signal-to-noise of the standard LGS algorithm up to $k_\mathrm{max}=0.4h/\mathrm{Mpc}$.  The EGS and LGS reconstruction algorithms are thus very similar in their ability to extract additional BAO information.  

With the goal of removing the entire second-order part of the nonlinear density, i.e.~nonlinear growth, shift and tidal terms, we introduced Eulerian and Lagrangian F2 algorithms.  The Eulerian version, EF2, performs almost as well as the growth-shift algorithms, missing only $6\%$ of the BAO signal-to-noise of the standard algorithm in our simulations.  The corresponding Lagrangian algorithm, LF2, performs somewhat worse, likely because small-scale modes are smoothed out too aggressively in this case.  Motivated by 2LPT results, we also considered random-random reconstructions, where random catalogs are shifted by the positive and negative Zeldovich displacement.  They turn out to perform significantly worse than the other algorithms.  In summary, the standard LGS algorithm performs best, but two new Eulerian reconstruction algorithms, EGS and EF2, perform almost equally well; see Fig.~\ref{fig:SixcumSNsq}.

For all our Eulerian algorithms, the reconstructed power spectrum is obtained by adding to the unreconstructed power spectrum simple model-independent 3-point and 4-point statistics of the unreconstructed density, like $\la\delta^2|\delta\ra$ or $\la\delta^2|\delta^2\ra$.  This makes it very transparent how these Eulerian reconstructions yield additional BAO information by exploiting higher-order N-point statistics. 

Modeling reconstructions in LPT and expanding all fields consistently up to $\mathcal{O}(\delta_0^2)$ shows that Eulerian and Lagrangian reconstructions change the density in exactly the same way (if a new correction term is included that arises by modeling the displacement field of clustered catalogs such that it is evaluated at Eulerian rather than Lagrangian locations, which matters beyond linear order).  We confirmed in simulations that density changes due to Eulerian and Lagrangian reconstructions look nearly the same in 2D density slices, see Fig.~\ref{fig:slices0pt01}.
Given that Eulerian and Lagrangian algorithms agree at the field level in theory and simulations and that they enhance BAO information at the power spectrum level in a very similar way, they can be regarded as Eulerian and Lagrangian incarnations of the same underlying reconstruction principles.  By making this connection one can argue that
 the standard Lagrangian LGS reconstruction is partially so successful because it automatically combines BAO information from specific well-motivated 2-, 3- and 4-point statistics of the unreconstructed density.  This provides a novel theoretical argument for the robustness and success of the standard LGS reconstruction algorithm.

Using specific splits of the mass density in simulations, we show that most of the additional BAO signal-to-noise from Eulerian reconstructions is due to 3-point statistics, while 4-point statistics add very little.  The 3-point contribution can be split further into parts of type $\la\delta_0|\delta_0^3\ra$ and $\la\delta_0^2|\delta_0^2\ra$, where $\delta_0$ is the linear density.  Our simulations show that the $13$ part of the 3-point contribution is responsible for sharpening the BAO wiggles, whereas the $22$ part leads to shifts and damping of the wiggles, in agreement with previous literature \cite{padmanabhan0812}.

The specific 3-point statistics that enter Eulerian reconstructions come in the form of cross-spectra of the unreconstructed density with the squared unreconstructed density, $\delta^2$, a shift term, $\vs\cdot\nabla\delta$, and a tidal term, $K^2$.  The same cross-spectra also arise when considering maximum-likelihood estimators for the amplitudes of the components of the tree-level DM bispectrum \cite{marcel1411}.  This suggests that the bispectrum of the unreconstructed  density does not contain much additional BAO information beyond that already recovered by reconstruction algorithms
(i.e.~the additional BAO information that the bispectrum could in principle yield is likely very correlated with the additional BAO information in the reconstructed power spectrum compared to the unreconstructed power spectrum).  This intuitive expectation should be tested more rigorously and quantitatively in the future.

\changed{An important limitation of the results presented here is that we restricted ourselves to dark matter in real space. We leave the important extension to galaxies in redshift space  for future work. Our findings can also be extended in many other directions. For example, it would be interesting}
 to study configuration space correlation functions in simulations and theory, and to include higher order modeling corrections. Ideas from previous studies \cite{SeoBAOShift,TassevRec,Achitouv2015} that aim at optimizing the standard reconstruction algorithm, e.g.~by improved weightings or iterative reconstructions, could also be applied to each of our new algorithms. Our results could also be useful for improving combinations of BAO analyses (fitting only for the BAO scale with templates) with redshift space distortion analyses (using the full shape of the redshift-space 2-point function), because the covariance between these analyses could be expressed in terms of covariances between 2-, 3- and 4-point functions that can be modeled perturbatively.  \changed{Ultimately it would  be  exciting to apply the new algorithms to real data.}

\section*{Acknowledgements}
We are grateful to Uro\v{s} Seljak for initial collaboration and many helpful discussions.  We also thank Nikhil Padmanabhan and Martin White for useful discussions, and Martin White for providing the RunPB TreePM N-body simulations that were used for parts of this paper.  This research used resources of the National Energy Research Scientific Computing Center (NERSC), a DOE Office of Science User Facility supported by the Office of Science of the U.S.~Department of Energy under Contract No.~DE-AC02-05CH11231.

\appendix

\section{Lagrangian Modeling}
\label{se:LagrangianTheory}

This appendix models Lagrangian reconstruction algorithms using the Lagrangian perspective of structure formation following \cite{padmanabhan0812}.  We start by modeling LGS reconstruction non-perturbatively in terms of the displacement field. To simplify this, we then review Lagrangian Perturbation Theory and use it to model LGS reconstruction perturbatively up to second order in the linear density. Finally, we model all other Lagrangian reconstruction algorithms mentioned in the main text. Throughout this appendix we include a second order correction term that arises when assuming that clustered catalogs are displaced from Eulerian rather than Lagrangian positions.

\subsection{Nonperturbative Lagrangian Modeling of Lagrangian Growth-Shift Reconstruction}

In the Lagrangian picture of structure formation, initial positions $\vq$ in Lagrangian space are displaced by $\vPsi(\vq)$ to obtain Eulerian positions $\vx$,
\begin{equation}
  \label{eq:6}
  \vx = \vq + \vPsi(\vq).
\end{equation}
This can be related to the mass overdensity by \cite{9604020}
\begin{equation}
  \label{eq:2}
  \delta^\theo(\vx) = \int\d^3 \vq\; \delta_D(\vx-\vq-\vPsi(\vq)) - 1,
\end{equation}
where $\delta_D$ is the Dirac delta and we use superscript `$\theo$' to denote theoretical mass densities.  The Fourier transform is
\begin{equation}
  \label{eq:delta_k_from_Psi_k}
  \delta^\theo(\vk) = \int \d^3 \vx\; e^{-i\vk\cdot\vx}\delta^\theo(\vx)
=\int \d^3 \vq\; e^{-i\vk\cdot\vq}\left(
e^{-i\vk\cdot\vPsi(\vq)}-1\right).
\end{equation}

The LGS-reconstructed density is obtained as follows. First, the negative Zeldovich displacement $\vs$ is calculated from the  non-linear density with \eqq{vs_def2}.   This is used to displace the original density, yielding the `displaced' field $\delta_d$ which can be modeled by
\begin{equation}
  \label{eq:delta_d_shifted}
  \delta_d^\theo(\vk) = \int\d^3\vq\;e^{-i\vk\cdot\vq}\left(e^{-i\vk\cdot [\vPsi(\vq)+\vs(\vx)]}-1\right).
\end{equation}
The theory displacement $\vPsi$ is evaluated at the initial Lagrangian position $\vq$, whereas the negative Zeldovich displacement $\vs$ obtained from the observations is evaluated at the final Eulerian location $\vx$. Instead of $\vs(\vx)$ previous literature uses $\vs(\vq)$ which misses a correction at second order in the fields (this correction is investigated further with simulations at the end of Appendix \ref{se:SliceComparisons}).  Next, a uniform or random sample of particles is displaced by the same vector $\vs$ evaluated at the uniform positions $\vq$ to get the `shifted' field $\delta_s$, which can be modeled by
\begin{equation}
  \label{eq:delta_s_shifted}
  \delta_s^\theo(\vk) = \int\d^3\vq\; e^{-i\vk\cdot\vq}\left(
e^{-i\vk\cdot\vs(\vq)}-1
\right).
\end{equation}
The difference of these two fields gives the reconstructed density for the standard LGS reconstruction algorithm,
\begin{equation}
  \label{eq:delta_LGS_def_appendix}
  \delta_\mathrm{LGS}^\rec \equiv \delta_d-\delta_s,
\end{equation}
which can be modeled by
\begin{equation}
\delta_\mathrm{LGS}^{\rec,\theo}(\vk)= \int\d^3\vq \;e^{-i\vk\cdot\vq}
\,
\left(
e^{-i\vk\cdot[ \vPsi(\vq)+\vs(\vx)]} - e^{-i\vk\cdot\vs(\vq)}
\right).
\end{equation}
This can be rearranged nonperturbatively to\footnote{This can be seen
by adding and
  subtracting $\delta^\theo$ 
given by \eqq{delta_k_from_Psi_k}, which yields
\begin{equation}
  \label{eq:delta_LGS_footnote}
  \delta_\mathrm{LGS}^{\rec,\theo}(\vk) 
= \delta^\theo(\vk) + \int\d^3\vq \;e^{-i\vk\cdot\vq}
\,
\left(
e^{-i\vk\cdot[ \vPsi(\vq)+\vs(\vx)]} - e^{-i\vk\cdot\vs(\vq)}
- e^{-i\vk\cdot\vPsi(\vq)}+1
\right).
  \end{equation}
}
  \begin{equation}
  \label{eq:delta_LGS_a_plus_b}
  \delta_\mathrm{LGS}^{\rec,\theo}
= \delta^\theo +
\delta_\mathrm{LGS}^{\rec,\mathrm{(a)}} + \delta_\mathrm{LGS}^{\rec,\mathrm{(b)}},
  \end{equation}
where $\delta^\theo$ models the nonlinear  density before reconstruction and we defined
\begin{equation}
  \label{eq:delta_LGS_a}
\delta_\mathrm{LGS}^{\rec,\mathrm{(a)}}(\vk) \equiv  \int\d^3\vq \;e^{-i\vk\cdot\vq}
\left(
e^{-i\vk\cdot[\vPsi(\vq)+\vs(\vq)]}
-e^{-i\vk\cdot\vPsi(\vq)}
-e^{-i\vk\cdot\vs(\vq)}
+1 \right). 
\end{equation}
An additional correction term (b) arises because the clustered catalog displacement
is evaluated at Eulerian positions, $\vs(\vx)$, instead of
Lagrangian positions, $\vs(\vq)$,
\begin{equation}
  \label{eq:delta_LGS_b}
\delta_\mathrm{LGS}^{\rec,\mathrm{(b)}}(\vk) \equiv
   \int\d^3\vq \;e^{-i\vk\cdot\vq}
e^{-i\vk\cdot\vPsi(\vq)}\left(
e^{-i\vk\cdot\vs(\vx)}-e^{-i\vk\cdot\vs(\vq)}
\right).
\end{equation}
Expressing the exponentials in \eqq{delta_LGS_a} as power series gives
\begin{equation}
  \label{eq:delta_LGS_a_sum}
  \delta_\mathrm{LGS}^{\rec,\mathrm{(a)}}(\vk) =  \int\d^3\vq \;e^{-i\vk\cdot\vq}
  \sum\limits_{n=2}^\infty
  \sum\limits_{m=1}^{n-1}
  \frac{(-i)^n}{n!}
  \binom{n}{m}
  \left[\vk\cdot\vPsi(\vq)\right]^m
  \left[\vk\cdot\vs(\vq)\right]^{n-m},
\end{equation}
where only terms mixing $\vPsi$ and $\vs$ survive ($m\ge 1$ and
$n-m\ge 1$).  Similarly, if we write 
$$\vs(\vx)=\vs(\vq)+\vH(\vq)$$
for some function $\vH$ (discussed in more detail later, see \eqq{vsx_vsq}), then \eqq{delta_LGS_b} becomes
\begin{equation}
  \label{eq:delta_LGS_b_sum}
    \delta_\mathrm{LGS}^{\rec,\mathrm{(b)}}(\vk) =  \int\d^3\vq
    \;e^{-i\vk\cdot\vq}
e^{-i\vk\cdot\vPsi(\vq)}
\sum\limits_{n=1}^\infty
\sum\limits_{m=1}^{n}
\frac{(-i)^n}{n!}
\binom{n}{m}
\left[\vk\cdot\vH(\vq)\right]^m
\left[\vk\cdot\vs(\vq)\right]^{n-m}.
\end{equation}

All expressions above are fully nonperturbative in the displacement field. They can be
simplified further in  Lagrangian Perturbation Theory
(LPT) which we briefly review in the next section.

\subsection{Review of Lagrangian Perturbation Theory (LPT)}
In LPT the displacement $\vPsi$ is
expanded perturbatively in powers of the linear density contrast
$\delta_0$,
\begin{equation}
  \label{eq:LPTexpansion}
  \vPsi = \vPsi^{(1)} + \vPsi^{(2)}+\cdots,
\end{equation}
where \cite{Bouchet9406013}
\begin{equation}
  \label{eq:33}
  \vPsi^{(n)}(\vk) = \frac{i}{n!}\int\prod\limits_{i=1}^n
\left[\frac{\d^3 \vk_i}{(2\pi)^3}\right] \,
(2\pi)^3
\delta_D(\vk_1+\cdots+\vk_n-\vk)
\,\vL^{(n)}(\vk_1, \dots, \vk_n; \vk)
\,\delta_0(\vk_1)\cdots\delta_0(\vk_n).
\end{equation}
The first and second order kernels are (see e.g.~\cite{BernardeauReview,2008PhRvD..77f3530M,1995MNRAS.276..115C,1996MNRAS.282..455C,rampf1203SPTvsLPTkernels})
\begin{eqnarray}
  \label{eq:34}
  \vL^{(1)}(\vk) &=& \frac{\vk}{k^2},\\
  \vL^{(2)}(\vk_1,\vk_2;\vk) &=&
  \frac{2}{7}\frac{\vk}{k^2}\left[
1-\mathsf{P}_2(\hat{\vk}_1\cdot\hat{\vk}_2)
\right],
\end{eqnarray}
where $\hat{\vk}\equiv \vk/k$ and $\mathsf{P}_2$ is the $l=2$ Legendre polynomial.

The density contrast \eq{delta_k_from_Psi_k} corresponding to this perturbative displacement field 
is
\begin{equation}
  \label{eq:36}
  \delta^\theo(\vk) = \delta_\mathrm{LPT}^{(1)}(\vk) + \delta_\mathrm{LPT}^{(2)}(\vk)+\cdots,
\end{equation}
where $\delta_\mathrm{LPT}^{(1)}=\delta_0$ and 
\begin{equation}
  \label{eq:delta2_LPT_1}
  \delta^{(2)}_\mathrm{LPT}(\vk) = \int\d^3\vq\, e^{-i\vk\cdot\vq}
\left[
-i\vk\cdot\vPsi^{(2)}(\vq) - \frac{1}{2}\left(
\vk\cdot\vPsi^{(1)}(\vq)
\right)^2
\right],
\end{equation}
noting that $\vPsi^{(n)}(\vq)$ are in Lagrangian configuration space. To simplify
the comparison with Eulerian SPT, we write the second order field in
terms of the kernel $F_2^\mathrm{LPT}$ defined by
\begin{equation}
  \label{eq:37}
  \delta_\mathrm{LPT}^{(2)}(\vk) = 
\int_{\vk_i}^*
F_2^\mathrm{LPT}(\vk_1,\vk_2)\,\delta_0(\vk_1)\delta_0(\vk_2).
\end{equation}
where we used the shorthand integral notation of \eqq{DefShortInt}.   The $F_2^\mathrm{LPT}$ kernel has components $F^\mathrm{LPT}_{2,11}$ from $\vPsi^{(1)}\vPsi^{(1)}$ in \eqq{delta2_LPT_1} and $F^\mathrm{LPT}_{2,2}$ from $\vPsi^{(2)}$, i.e.~
\begin{equation}
  \label{eq:39}
  F_2^\mathrm{LPT} = F_{2,11}^\mathrm{LPT} + F_{2,2}^\mathrm{LPT}
\end{equation}
with\footnote{Note that for $\vk=\vk_1+\vk_2$ we have
  \begin{equation}
    \label{eq:kL1_kL1_simplified}
\left[\vk\cdot\vL^{(1)}(\vk_1)\right]
\left[\vk\cdot\vL^{(1)}(\vk_2)\right]
=
\frac{4}{3} + \left(\frac{k_1}{k_2}+\frac{k_2}{k_1}\right)
\hat{\vk}_1\cdot\hat{\vk}_2
+ \frac{2}{3}\mathsf{P}_2(\hat{\vk}_1\cdot\hat{\vk}_2).    
  \end{equation}
Also note that since $F_{2,11}^\mathrm{LPT}(\vk_1,-\vk_1)=0$ and
$F_{2,2}^\mathrm{LPT}(\vk_1,-\vk_1)=0$ the
 corresponding contributions to 
$\delta^{(2)}_\mathrm{LPT}$ have zero spatial average,
$\langle\delta^{(2,11)}_\mathrm{LPT}(\vx)\rangle=0$  
and
$\langle\delta^{(2,2)}_\mathrm{LPT}(\vx)\rangle=0$.
}
\begin{eqnarray}
  \label{eq:F2_11_LPT}
  F_{2,11}^\mathrm{LPT}(\vk_1,\vk_2) &=&\frac{2}{3}
+ \frac{1}{2}\,\left(\frac{k_1}{k_2}+\frac{k_2}{k_1}\right)
\hat{\vk}_1\cdot\hat{\vk}_2 +
\frac{1}{3}\mathsf{P}_2(\hat{\vk}_1\cdot\hat{\vk}_2),\\
  \label{eq:F2_2_LPT}
  F_{2,2}^\mathrm{LPT}(\vk_1,\vk_2) &=&\frac{1}{7} - \frac{1}{7}
\mathsf{P}_2(\hat{\vk}_1\cdot\hat{\vk}_2).
\end{eqnarray}
This agrees with the $F_2$ kernel from Eulerian SPT
\cite{rampf1203SPTvsLPTkernels,white1504.03677},
\begin{equation}
  \label{eq:41}
F_2^\mathrm{LPT}(\vk_1,\vk_2)=F_2^\mathrm{SPT}(\vk_1,\vk_2) =   
\frac{17}{21}
+ \frac{1}{2}\,\left(\frac{k_1}{k_2}+\frac{k_2}{k_1}\right)
\hat{\vk}_1\cdot\hat{\vk}_2 +
\frac{4}{21}\mathsf{P}_2(\hat{\vk}_1\cdot\hat{\vk}_2).
\end{equation}
The $F^\mathrm{LPT}_{2,2}$ contribution from $\vL^{(2)}$ does not change the shift term, but it changes the amplitude of the growth term by $21\%$ and that of the tidal contribution by $43\%$.

\subsection{LGS Reconstruction in 2LPT}

Working in LPT and keeping terms up to second order in the linear density $\delta_0$,
the contribution (a) of \eqq{delta_LGS_a_sum} to the reconstructed density becomes
\begin{equation}
  \label{eq:42}
  \delta_\mathrm{LGS}^{\rec,\mathrm{(a)}}(\vk) = 
- 
\int_{\vk_i}^*
\left[\vk\cdot\vL^{(1)}(\vk_1)\right]
\left[\vk\cdot\vL^{(1)}(\vk_2)\right]
W_R(k_2)\delta_0(\vk_1)\delta_0(\vk_2),
\end{equation}
where we wrote $\vPsi(\vq)$ and $\vs(\vq)$ in Fourier
space in terms of $\vL^{(1)}$. From 
Eqs.~\eq{kL1_kL1_simplified} and \eq{F2_11_LPT} we get
\cite{padmanabhan0812}
\begin{equation}
  \label{eq:delta_LGS_a_F2_11}
  \delta_\mathrm{LGS}^{\rec,\mathrm{(a)}}(\vk) = 
- 
\int_{\vk_i}^*
2F^\mathrm{LPT}_{2,11}(\vk_1,\vk_2)
W_R(k_2)\delta_0(\vk_1)\delta_0(\vk_2).
\end{equation}

Next, we calculate at second order in LPT the correction term (b) given by \eqq{delta_LGS_b_sum} (coming from evaluating clustered catalog displacements at Eulerian rather than Lagrangian locations, see also Appendix~\ref{se:SliceComparisons}).  At second order  we have 
\begin{equation}
  \label{eq:vsx_vsq}
 \vs(\vx)\approx \vs(\vq)+(\vPsi(\vq)\cdot\nabla)\vs(\vq),
\end{equation}
so that \eqq{delta_LGS_b_sum} becomes at second order
\begin{equation}
  \label{eq:delta_LGS_b2}
\delta_\mathrm{LGS}^{\rec,\mathrm{(b)}}(\vk)  \approx
- i \int\d^3\vq\; e^{-i\vk\cdot\vq} e^{-i\vk\cdot\vs(\vq)}
\vk\cdot (\vPsi(\vq)\cdot\nabla)\vs(\vq).
\end{equation}
Then
\begin{eqnarray}
  \label{eq:delta_LGS_b_v2}
  \delta_\mathrm{LGS}^{\rec,\mathrm{(b)}}(\vk) &=&
\int_{\vk_i}^*
\left(\vk\cdot\vL^{(1)}(\vk_2)\right)
\left(\vk_2\cdot\vL^{(1)}(\vk_1)\right) W_R(k_2) 
\delta_0(\vk_1)\delta_0(\vk_2)\\
&=&
  \label{eq:delta_LGS_b_v3}
\int_{\vk_i}^*
\left[ \frac{1}{3}+\frac{k_2}{k_1}\hat{\vk}_1\cdot\hat{\vk}_2
+\frac{2}{3}\mathsf{P}_2(\hat{\vk}_1\cdot\hat{\vk}_2) \right]
W_R(k_2) 
\delta_0(\vk_1)\delta_0(\vk_2).
\end{eqnarray}
Adding Eq.~\eq{delta_LGS_b_v3} to \eq{delta_LGS_a_F2_11} cancels the tidal part and one of the two shift terms of
$2F^\mathrm{LPT}_{2,11}$, so that the model for the total
reconstructed density becomes simply
\begin{equation}
  \label{eq:delta_LGS_2nd_order_appendix}
  \delta_\mathrm{LGS}^{\rec,\theo}(\vk) = \delta^\theo(\vk) - 
\int_{\vk_i}^*
\left[1+\frac{k_1}{k_2}\hat{\vk}_1\cdot\hat{\vk}_2\right]
W_R(k_2)\delta_0(\vk_1)\delta_0(\vk_2),
\end{equation}
which agrees with \eqq{delta_LGS_2nd_order} in the main text and corresponds to \eqq{delta_LGS_x_LPTfinal} in configuration space.
Thus, in this LPT picture, standard LGS reconstruction subtracts growth and shift terms from the nonlinear density $\delta$.

The second-order contribution to the density change caused by reconstruction came from
terms with kernels of the type $(L^{(1)})^2$ but no $L^{(2)}$ kernel
contributed, which means that this reconstruction cannot fully reverse
second-order non-linearities \cite{padmanabhan0812}. We can generalize
this statement using \eqq{delta_LGS_a_sum}: At $n$-th order LPT, part
(a) of the reconstruction does not have an explicit contribution from
the kernel $L^{(n)}$, but only products of lower-order kernels
$L^{(r_1)}\cdots L^{(r_s)}$ (with $s\ge 2$ and $\sum_i r_i=n$).  For
example, at third order only kernels of the type $(L^{(1)})^3$ and
$(L^{(1)})^2L^{(2)}$ contribute but not $L^{(3)}$.  The same statement
is true for the (b) part of the reconstruction given by
\eqq{delta_LGS_b_sum}.\footnote{The only possibility where a summand
  in \eqq{delta_LGS_b_sum} is not a product of at least two fields is
  if $m=n=1$, in which case there is a contribution going like
  $\vH^{(n)}$.  However, Taylor expanding
  $\vH(\vq)=\vs(\vq+\vPsi)-\vs(\vq)$ in $\vPsi$
  implies that this is always a product of at least two fields that depend on $L$ kernels
  (schematically, suppressing derivatives and vectors, $H \sim \Psi s
  + \tfrac{1}{2}\Psi^2 s +\tfrac{1}{6}\Psi^3 s+\cdots$). }
Consequently,  the standard LGS reconstruction cannot fully remove any of the `pure' $L^{(n)}$
contributions to the nonlinear density, unless products of lower-order
kernels conspire to agree with $L^{(n)}$.  It can of course still improve BAO information by removing `non-pure' non-linearities, e.g.~of type $(L^{(1)})^2$, and this turns out to be very successful in practice.

\subsection{2LPT Models for 6 Lagrangian Reconstructions}
\label{se:alt_rec_2nd_order}
We introduced 6 possible Lagrangian reconstruction algorithms in Section~\ref{se:LagrangianRecs}.  At first order, they all agree with the linear density, but at higher order they differ from each other.  This appendix models these algorithms in 2LPT.

The four relevant densities for the reconstructions are clustered catalogs displaced by $\vs$ or $-\vs$, denoted by $\delta_d[\vs]$ and $\delta_d[-\vs]$, and random catalogs shifted by $\vs$ or $-\vs$, denoted by $\delta_s[\vs]$ and $\delta_s[-\vs]$.  We model them in 2LPT by expanding Eqs.~\eq{delta_d_shifted} and \eq{delta_s_shifted} consistently at $\mathcal{O}(\delta_0^2)$.  For clustered catalogs, the second order part gives
\begin{equation}
  \label{eq:83}
  \delta^{(2)}_d[\pm\vs](\vk) = \int_{\vk_i}^*\Big\{
\left[1\mp W_R(k)\right]F_2(\vk_1,\vk_2) \mp W_R(k_2) \left[1+\frac{k_1}{k_2}\hat{\vk}_1\cdot\hat{\vk}_2\right] + W_R(k_1)W_R(k_2)F^\mathrm{LPT}_{2,11}(\vk_1,\vk_2)
\Big\}\delta_0(\vk_1)\delta_0(\vk_2).
\end{equation}
Here we used $\vs(\vx)=\vs(\vq)+\vH(\vq)$, where $\vH^{(2)}(\vq)=(\vPsi^{(1)}(\vq)\cdot\nabla)\vs^{(1)}(\vq)$ and $-i\vk\cdot\vH^{(2)}(\vk)$ is given by \eqq{delta_LGS_b_v3}.  Also, note that the negative Zeldovich displacement $\vs(\vq)$ computed from the clustered catalog is
\begin{equation}
  \label{eq:82}
  \vs^{(2)}(\vk) = -i \vL^{(1)}(\vk) W_R(k)\int_{\vk_i}^* F_2(\vk_1,\vk_2)\delta_0(\vk_1)\delta_0(\vk_2)
\end{equation}
at second order.  For the shifted randoms, the second order part is
\begin{equation}
  \label{eq:84}
  \delta_s^{(2)}[\pm\vs](\vk) = \int_{\vk_i}^*\Big\{
\mp W_R(k)F_2(\vk_1,\vk_2) + W_R(k_1)W_R(k_2)F^\mathrm{LPT}_{2,11}(\vk_1,\vk_2)
\Big\}\delta_0(\vk_1)\delta_0(\vk_2).
\end{equation}
The reconstructed densities of
Eqs.~\eq{rec_combi1}-\eq{rec_combi6} up to second order are then
\begin{align}
  \label{eq:delta_d_s_minus_delta_s_s}
 \delta_d[\vs]-\delta_s[\vs]   &=\delta_0(\vk) + 
\int^*_{\vk_i}\left\{
F_2(\vk_1,\vk_2) - W_R(k_2)\left[1+\tfrac{k_1}{k_2}\mu\right]
\right\}\delta_0(\vk_1)\delta_0(\vk_2),\\
\nonumber
 \tfrac{1}{2}\left\{\delta_d[\vs]+\delta_d[-\vs]\right\}   &=\delta_0(\vk)
 + 
\int^*_{\vk_i}\Big\{
F_2(\vk_1,\vk_2) +
W_R(k_1)W_R(k_2)F_{2,11}^\mathrm{LPT}(\vk_1,\vk_2)
\Big\}\delta_0(\vk_1)\delta_0(\vk_2),
\\
\nonumber
 \delta_d[\vs]+\delta_s[-\vs]   &=\delta_0(\vk) +
\int^*_{\vk_i}\Big\{
F_2(\vk_1,\vk_2) +2 W_R(k_1)W_R(k_2)F_{2,11}^\mathrm{LPT}(\vk_1,\vk_2)
-W_R(k_2)\left[1+\tfrac{k_1}{k_2}\mu
\right]
\Big\}\delta_0(\vk_1)\delta_0(\vk_2),\qquad\;
  \\
\nonumber
 \delta_s[\vs]+\delta_d[-\vs]   &=\delta_0(\vk) + 
\int^*_{\vk_i}\Big\{
F_2(\vk_1,\vk_2) +2
W_R(k_1)W_R(k_2)F_{2,11}^\mathrm{LPT}(\vk_1,\vk_2)
+W_R(k_2)\left[1 + \tfrac{k_1}{k_2}\mu
\right]
\Big\}\delta_0(\vk_1)\delta_0(\vk_2),
\\
\label{eq:delta_s_s_minus_delta_s_minus_s}
\delta - c\left\{ \delta_s[\vs]+\delta_s[-\vs]\right\}   &=\delta_0(\vk) +
\int^*_{\vk_i}\left\{
F_2(\vk_1,\vk_2) -2\,c\, W_R(k_1)W_R(k_2)F_{2,11}^\mathrm{LPT}(\vk_1,\vk_2)
\right\}\delta_0(\vk_1)\delta_0(\vk_2),
\\
\nonumber
 \delta_d[-\vs]-\delta_s[-\vs]   &=\delta_0(\vk) + 
\int^*_{\vk_i}\Big\{
F_2(\vk_1,\vk_2) 
+W_R(k_2)\left[1 + \tfrac{k_1}{k_2}\mu
\right]
\Big\}\delta_0(\vk_1)\delta_0(\vk_2), 
\end{align}
where $\mu\equiv \hat{\vk}_1\cdot\hat{\vk}_2$.
From the six possibilities, only combinations \eq{delta_d_s_minus_delta_s_s} and
\eq{delta_s_s_minus_delta_s_minus_s} suppress nonlinear growth and shift.

\section{Slice Comparisons}
\label{se:SliceComparisons}
\begin{figure}[p]
\centerline{
\includegraphics[width=0.9\textwidth]{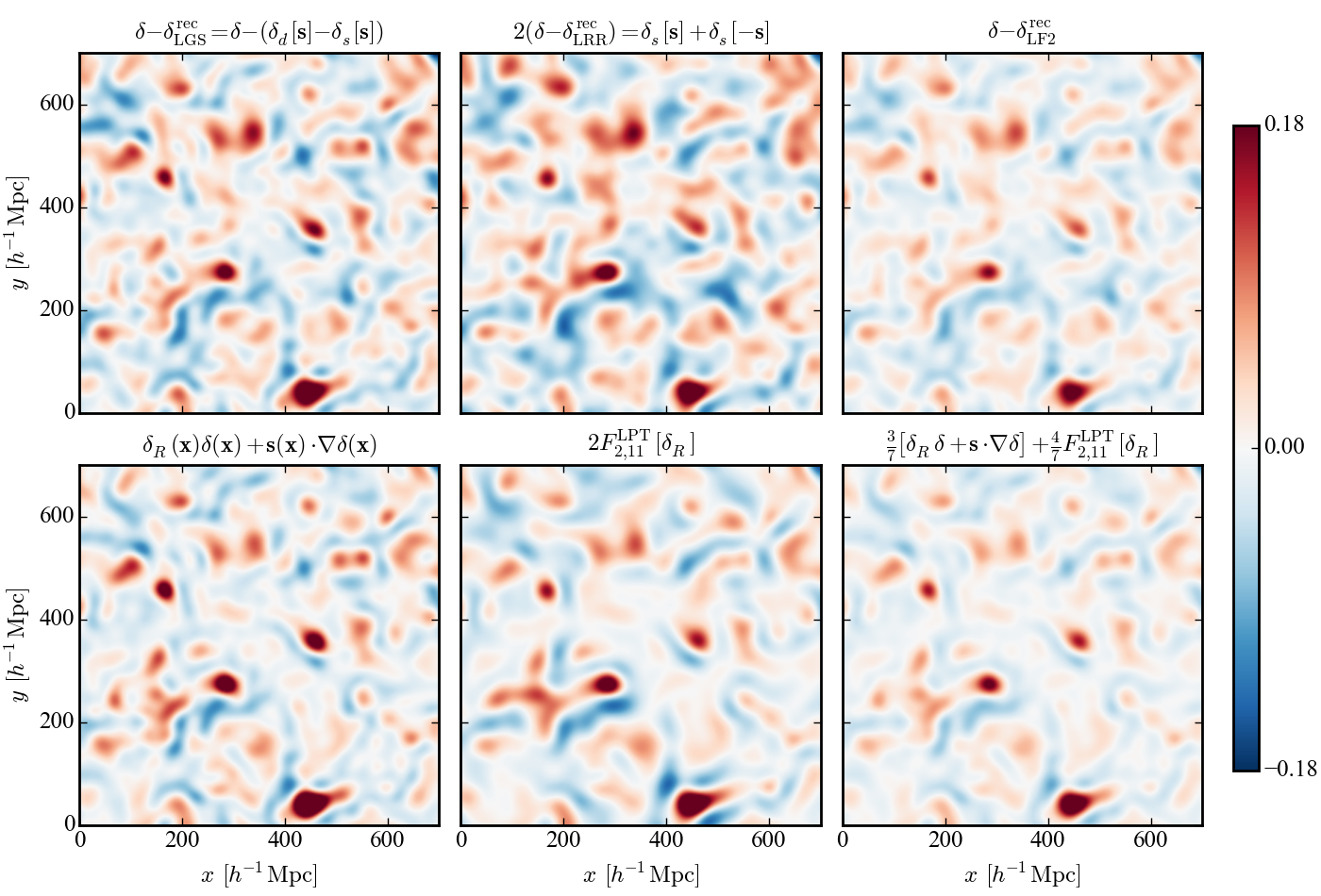}}
\caption{Comparison of Lagrangian (top) and Eulerian (bottom) reconstructions at the field level for growth-shift (left), random-random (middle) and F2 (right) algorithms.  The panels show 2D slices of excess densities $\delta-\delta^\mathrm{rec}$ that are removed from the density when performing reconstruction. 
The excess densities in the bottom panels are quadratic in the nonlinear density measured from simulations.  They coincide with the excess densities expected from modeling the Lagrangian reconstructions in the upper panels with 2LPT, if the linear densities in Eqs.~\eq{delta_LGS_x_LPTfinal}, \eq{delta_LRR_x_LPTfinalMainText} and \eq{delta_LF2_2nd_order} are replaced by the nonlinear ones.
For LRR and ERR, twice the excess density is shown to enhance color contrast.
   We use $R=15h^{-1}\mathrm{Mpc}$ Gaussian smoothing for the reconstructions.  To highlight large scales, each final field is additionally smoothed externally with a $R=15h^{-1}\mathrm{Mpc}$ Gaussian. The clustered catalog is obtained from a $1\%$ random subsample of one RunPB N-body realization; the random catalog from equally many randomly distributed particles. 
}
\label{fig:slices0pt01}
\end{figure}

\begin{figure}[p]
\centerline{
\includegraphics[width=0.9\textwidth]{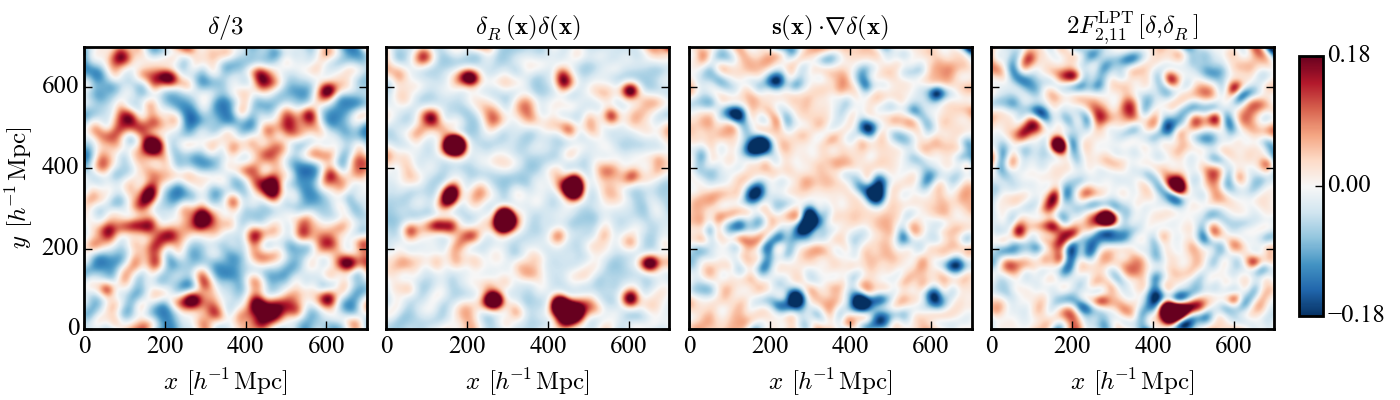}
}
\caption{\emph{Left panel:} Nonlinear density before reconstruction (divided by $3$ for better visibility).   \emph{Middle panels:} Individual growth and shift contributions to the Eulerian growth-shift reconstruction.  \emph{Right panel:}  Expected excess density for growth-shift reconstruction if the correction term (b) in \eqq{delta_LGS_a_plus_b} is ignored.  The plot shows term (a) given by \eqq{delta_LGS_a_F2_11} (multiplied by -1). All densities were obtained   from the same $1\%$  subsample as in Fig.~\ref{fig:slices0pt01} and are plotted in the same way.}
\label{fig:slices_contris}
\end{figure}

To compare reconstruction algorithms at the field level and as an illustrative sanity check of their 2LPT models, we write 
\begin{equation}
  \label{eq:85}
  \delta_\mathrm{rec}=\delta-(\delta-\delta_\mathrm{rec}).
\end{equation}
Here, $\delta-\delta_\mathrm{rec}$ is the nonlinear excess density that is removed from the full nonlinear density when applying reconstruction.   Fig.~\ref{fig:slices0pt01} shows 2D slices of this excess density for the three Lagrangian reconstruction algorithms (top) in comparison with the three corresponding Eulerian algorithms (bottom).  For each column of the plot, upper and lower panels are rather similar, i.e~Eulerian and Lagrangian reconstructions change the density in a very similar way.  This shows qualitatively that Eulerian and Lagrangian reconstructions agree with each other at the field level on the large scales that are highlighted by these slice plots.  The agreement between upper and lower panels in Fig.~\ref{fig:slices0pt01} also illustrates that second order LPT provides an accurate description of the reconstruction algorithms on large scales.   The different Lagrangian reconstructions shown in the different columns are  also rather correlated, but they differ somewhat, which is expected since the 2LPT models of these methods differ from each other.

Fig.~\ref{fig:slices_contris} shows some additional slice plots for reference and comparison. The density before reconstruction is a few times larger than typical changes due to reconstruction.  The growth term $\delta_R\delta$ peaks at density peaks and troughs which are amplified by squaring the field. The magnitude of the shift term $\vs\cdot\nabla\delta$ is maximal at locations where the large-scale Zeldovich displacement is aligned with the density gradient, which is most pronounced near density peaks.  Since growth and shift terms have opposite signs near density peaks, their combination cancels the largest peaks to leave a smoother combined field (see Fig.~\ref{fig:slices0pt01}).

Finally, we check if displaced clustered catalogs should be modeled by evaluating the displacement field $\vs$ at Eulerian positions $\vx$ rather than Lagrangian positions $\vq$, which leads to the correction term (b) in \eqq{delta_LGS_a_plus_b}.  The panel on the right in Fig.~\ref{fig:slices_contris}  shows the expected excess density for LGS reconstruction if that correction term is ignored and only term (a) is used in \eqq{delta_LGS_a_plus_b}.  The same plot including the correction term (b) is shown in the bottom left panel of Fig.~\ref{fig:slices0pt01}.  These two panels should be compared with the top left panel of Fig.~\ref{fig:slices0pt01} which shows the true excess density for the full LGS reconstruction.  The impact of term (b) is not large but it does significantly smoothen some of the peaks and troughs of the density (slightly difficult to see in the slices because colors are truncated) and leads to slightly better agreement with the excess density for the full LGS scheme.  The slice plots thus suggest that the correction (b) should be included in the 2LPT modeling, although its effect is not large.  

It should also be noted that the Eulerian growth-shift reconstruction derived at the beginning of the paper from the continuity equation only agrees with the standard Lagrangian reconstruction in 2LPT if the correction (b) is included in the modeling of the latter.  An Eulerian reconstruction scheme that would only resemble term (a) of \eqq{delta_LGS_a_plus_b} yields a cumulative BAO signal-to-noise-squared of $\sim 150$ for $k_\mathrm{max}=0.4h/\mathrm{Mpc}$ in our simulations. This is significantly less than that for the EGS method which includes the correction term (b) and resembles the BAO information of LGS reconstruction more closely.  We take this as additional evidence that term (b) should be included.

\section{Newton-Raphson Method}
\label{se:NewtonRaphson}

We introduced Eulerian EF2 reconstruction in Section~\ref{se:EF2Rec} by subtracting from the nonlinear density the second order $F_2$ part, which was approximated to be quadratic in the observed rather than linear density.  Here we take a more formal approach to motivate this algorithm using the Newton-Raphson iteration to find a linear density that generates a given observed nonlinear density (at second order).  Note that our approach differs from the (more complicated) iteration scheme proposed in \cite{TassevRec},  which is based on maximizing the likelihood of the displacement field under certain assumptions.

In Eulerian Standard Perturbation Theory (SPT), the theoretical nonlinear DM density $\delta^\theo_\mathrm{NL}$ is modeled in terms of the linear density $\delta_0$ as
\begin{equation}
  \label{eq:deltaNL}
  \delta_\mathrm{NL}^\theo(\vx) = 
\delta_0(\vx) +
  F_2[\delta_0](\vx) + \cdots,
\end{equation}
where the second-order perturbation is
\begin{equation}
  \label{eq:7}
  F_2[\delta_0](\vx) = \frac{17}{21} \delta_0^{2}(\vx) - \vPsi_0(\vx)\cdot\nabla\delta_0(\vx) + \frac{4}{21}K_0^{2}(\vx),
\end{equation}
and the linear displacement $\vPsi_0$ and linear tidal term are defined in terms of $\delta_0$ by Eqs.~\eq{vPsi0}-\eq{tidal_x_def}. We neglect higher-order terms like $F_3[\delta_0](\vx)$ in \eqq{deltaNL}.   Given a fixed observed non-linear DM density $\delta_\mathrm{obs}$, our goal is to estimate the linear density $\delta_0$ that generated this observed density, i.e.~we want to find $\delta_0$ such that
\begin{equation}
  \label{eq:13}
f[\delta_0] \equiv \delta_0 + F_2[\delta_0] - \delta_\rm{obs} =0,
 \end{equation}
i.e.~we want to find roots of the functional $f$ with respect to $\delta_0$. This can be
achieved with the Newton-Raphson iteration method, where the
$(N+1)$-th iteration step is
\begin{equation}
  \label{eq:NRStep}
  \delta_0^{(N+1)}(\vk) = \delta_0^{(N)}(\vk) -
\int\frac{\d^3 \vk'}{(2\pi)^3}  \left(f'[\delta_0^{(N)}]\right)^{-1}_{\vk,\vk'} \,f[\delta_0^{(N)}]_{\vk'},
\end{equation}
where superscripts label iteration steps throughout this section.
$f'$ in \eqq{NRStep} is the functional derivative of $f$. This is a linear operator that can be written in matrix form as
\begin{equation}
  \label{eq:17}
f'[\delta_0]_{\vk,\vk'} =  \frac{\d f[\delta_0(\vk)]}{\d \delta_0(\vk')}
=
\delta_\rm{D}(\vk-\vk') + 2F_2(\vk',\vk-\vk')\delta_0(\vk-\vk').
\end{equation}
For $\delta_0\ll 1$, the inverse of the derivative is then
\begin{equation}
  \label{eq:16}
  \left(f'[\delta_0]\right)^{-1}_{\vk,\vk'} = \delta_\rm{D}(\vk-\vk') - 2
  F_2(\vk',\vk-\vk') \delta_0(\vk-\vk') + \mathcal{O}(\delta_0^2).
\end{equation}
The Newton-Raphson iteration scheme thus becomes
\begin{equation}
  \label{eq:78}
  \delta_0^{(N+1)}(\vk) = \delta_\mathrm{obs}(\vk) - F_2[\delta_0^{(N)}](\vk) 
+2\int\frac{\d^3 \vk'}{(2\pi)^3} F_2(\vk',\vk-\vk')\delta_0^{(N)}(\vk-\vk')
\left[\delta_0^{(N)}(\vk')-\delta_\mathrm{obs}(\vk')\right]
+ \mathcal{O}((\delta_0^{(N)})^3).
\end{equation}

If we start the iteration with $\delta^{(0)}_0=0$, then the first iteration step gives $\delta^{(1)}_0=\delta_\rm{obs}$,
i.e.~the linear density is approximated by the non-linear observed density.\footnote{We could as well start the iteration directly with $\delta_0^{(0)}=\delta_\mathrm{obs}$; then the N-th step would agree with the (N+1)-th step of the iteration started with $\delta_0^{(0)}=0$. } The second
iteration step gives
\begin{eqnarray}
  \label{eq:15}
  \delta^{(2)}_0(\vx) &=& \delta_\rm{obs} - F_2[\delta_\rm{obs}] +
  \mathcal{O}(\delta_\rm{obs}^3)\\
&=& \delta_\rm{obs}(\vx) - \frac{17}{21}\delta_\rm{obs}^2(\vx) 
+ \vPsi_\rm{obs}(\vx) \cdot\nabla\delta_\rm{obs}(\vx) -
\frac{4}{21}K_\rm{obs}^2(\vx) +
  \mathcal{O}(\delta_\rm{obs}^3)
\end{eqnarray}
This agrees with the EF2 algorithm from Section~\ref{se:EF2Rec} (up to  smoothing, which we ignored here for simplicity).

\section{Numerical Analysis Details}

\subsection{Simulation Convergence Tests}
\label{se:SimConvergence}
To check convergence of the \yucode simulations for the quantities and plots relevant to this paper, we perform several convergence tests.   First, we generate a single \yucode realization that matches the cosmology and phases of the initial conditions of one of the TreePM RunPB simulations.  We then measure all auto- and cross-spectra between $\delta$, $\delta^2$, $\vPsi\cdot\nabla\delta$ and $K^2$, corresponding to 2-, 3- and 4-point statistics required for Eulerian reconstructions.  All spectra from \yucode are slightly lower than the corresponding TreePM RunPB spectra over the whole $k$ range, and the deviations are smallest at low $k$.  The density power spectrum (2-point) is less than $0.8\%$ low over the whole $k$ range, 3-point spectra like $\la\delta^2|\delta\ra$ differ by less than $1\%$ at $k\leq 0.2h/\mathrm{Mpc}$ and by at most $1.5\%$ at higher $k$.  4-point spectra like $\la\delta^2|\delta^2\ra$ differ by less than $1.5\%$ at $k\leq 0.2h/\mathrm{Mpc}$ and by at most $2\%$ at higher $k$.  This is true for full DM samples as well as $1\%$ subsamples (with the same particles selected from the RunPB and \yucode simulation). In summary, \yucode simulations and TreePM RunPB simulations agree at the $2\%$-level.

Since BAO wiggles are only percent-level fluctuations on top of the broadband shape of the power spectrum, even differences at the $2\%$ level could potentially be problematic for BAO studies.  We expect however that broadband systematic inaccuracies in the simulations should cancel out when forming differences between wiggle and nowiggle simulations, as long as these systematics are present in wiggle and nowiggle simulations, which is a reasonably assumption.  To test this more quantitatively we vary the accuracy settings of the \yucode code and check if any results depend on them: First, we run the same simulations with only $20$ time steps and starting redshift $z_i=9$ (instead of $80$ time steps and $z_i=99$ used for the fiducial simulations).  Second, we also run them with $80$ time steps and $z_i=9$.  We generated every plot of this paper based on \yucode simulations for the three different accuracy settings.  All plots agree almost perfectly between the three settings, and changes are so small that they cannot be seen by eye. Therefore, our results do not depend on the accuracy settings of the \yucode code.  This provides additional evidence that the simulations are converged for the purposes of this paper.  For all plots that require nowiggle simulations,  we will only show plots for the fiducial \yucode simulations with $80$ time steps and $z_i=99$ because they should be most accurate.  For plots that do not require nowiggle simulations we use RunPB TreePM simulations instead.

\subsection{Dependence on Reconstruction Smoothing Scale}
\label{se:VaryR}
\begin{figure}[tp]
\centerline{
\includegraphics[height=0.25\textheight]{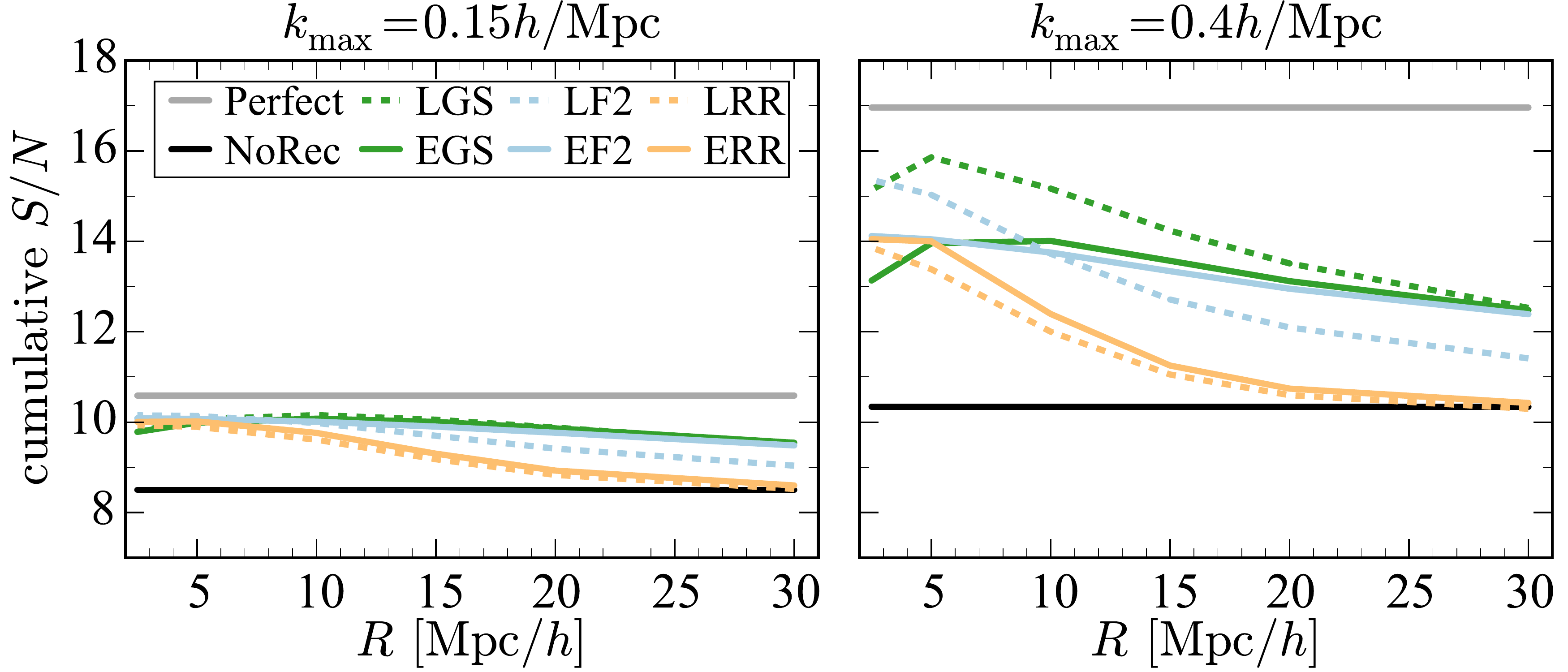}}
\caption{Cumulative BAO signal-to-noise for varying Gaussian smoothing scale $R$, using scales up to $k_\mathrm{max}=0.15h/\mathrm{Mpc}$ (left) or $k_\mathrm{max}=0.4h/\mathrm{Mpc}$ (right). Note that the shot noise in our setup is unrealistically small, $\bar{n}^{-1}=31\mathrm{Mpc}^3/h^3$. }
\label{fig:VarySmoothing}
\end{figure}

All numerical results in the main text assumed a fiducial Gaussian smoothing scale of $R=15\mathrm{Mpc}/h$ for the reconstruction algorithms, which was found to be optimal for the analysis in \cite{Padmanabhan2012BAORec,AndersonBAODR9_1203}.    Here we briefly discuss how our results depend on this choice.  Fig.~\ref{fig:VarySmoothing} shows the cumulative BAO signal-to-noise for different smoothing scales $R$.  For $k_\mathrm{max}=0.15h/\mathrm{Mpc}$, Eulerian and Lagrangian reconstructions perform very similarly independent of smoothing scale, which is expected because the 2LPT models of the algorithms agree.  If smaller scales are included, $k_\mathrm{max}=0.4\mathrm{Mpc}/h$, Eulerian and Lagrangian reconstructions still perform similarly for large smoothing scale, $R\gtrsim 15\mathrm{Mpc}/h$. In particular, EGS, EF2 and LGS all have the same performance for $R=30\mathrm{Mpc}/h$.  Towards smaller smoothing scales, especially for $R\sim 5\mathrm{Mpc}/h$, the LGS and LF2 algorithms improve significantly, whereas their Eulerian counterparts EGS and EF2 do not improve as much.  This shows the limitations of the Eulerian methods on small scales $k>0.15h/\mathrm{Mpc}$ and small smoothing scales $R\sim 5\mathrm{Mpc}/h$, where Lagrangian algorithms seem to work somewhat better; EGS misses up to $13\%$ of the total BAO signal-to-noise of LGS in the most extreme case (for $R=2.5\mathrm{Mpc}/h$ and $k_\mathrm{max}=0.4h/\mathrm{Mpc}$). However, it is important to note that the shot noise of our $1\%$ DM subsamples is unrealistically small, $\bar{n}^{-1}=31\mathrm{Mpc}^3/h^3$.  For more realistic, higher shot noise levels, the shot noise power starts to dominate on larger scales so that reconstruction performance is expected to be optimal for $R\sim 15\mathrm{Mpc}/h$ and decrease for smaller smoothing scales \cite{Padmanabhan2012BAORec}.  We also checked that our results do not qualitatively change if the Gaussian smoothing kernel is replaced by a steeper logistic smoothing function $W_{R;S}(k)=1/\{1+\exp[S(k-1/R)]\}$, where we tried $R=5\mathrm{Mpc}/h$ and $R=10\mathrm{Mpc}/h$ with $S=100$.

\subsection{Gaussian Covariance}
\label{se:GaussianCov}
Throughout this paper we assumed diagonal Gaussian covariances given by \eqq{GaussianNoise}.  We briefly argue here why this is a reasonable approximation for the purposes of estimating the BAO signal-to-noise. 

The non-Gaussian off-diagonal corrections to the power spectrum covariance that become important in the nonlinear regime can be modeled by \cite{IrshadUros1407}
\begin{equation}
  \label{eq:covNG}
  \mathrm{cov}(\hat P(k), \hat P(k')) = \la\hat P(k)\ra\la\hat P(k')\ra\left[
\delta_{k,k'}\frac{2}{N_\mathrm{modes}(k)}
+ c_\mathrm{NG}\right],
\end{equation}
where the non-Gaussian correction $c_\mathrm{NG}$ is independent of $k$ and $k'$.  It arises from the non-Gaussian coupling of large and small scale modes that leads to a broadband up- or down-shift of the small-scale power spectrum depending on the specific realization of large-scale modes in any given realization (see e.g.~\cite{Tobias1504.04366}). In other words, in some realizations the ratio of wiggle over nowiggle power spectra will be high over the whole $k$ range, while in other realizations that ratio will be low over the whole $k$-range if large-scale modes happen to be different (we confirmed this effect in our simulations).  In any given realization, the broadband rescaling of the power spectrum will be absorbed by the broadband part when fitting for BAO wiggles on top of some generic broadband function (as done in practice).  Therefore the non-Gaussian correction $c_\mathrm{NG}$ in \eqq{covNG} does not affect the BAO signal-to-noise and can be ignored for the purposes of this paper.   

We confirmed in simulations that the variance of spectra estimated from the mode-to-mode scatter in any single realization is in excellent agreement with the Gaussian expectation.  The variance estimated from the scatter between our \yucode realizations 1 and 2 also agrees with this, but it is significantly larger when estimated from the scatter between realizations 1 and 3 or 2 and 3. The reason is that our realization 3 has a higher overall amplitude over the whole $k$ range than the other two realizations, which is compatible with the arguments above and does not affect the significance of BAO wiggles in any given realization.

\bibliography{marcel_rec}

\end{document}